\documentclass{article}

\usepackage{lineno}
\usepackage{amsmath}
\usepackage{float}
\usepackage{comment}
\usepackage{graphicx}
\usepackage{makecell}
\usepackage{hyperref}
\usepackage[table]{xcolor}
\usepackage{color, colortbl}
\usepackage[percent]{overpic}
\usepackage{amsfonts}
\usepackage{multirow}
\usepackage{mathtools}
\usepackage{setspace}
\usepackage[]{algorithm2e}
\usepackage{setspace,caption}

\usepackage{authblk}

\title{Accelerating engineering design by automatic selection of simulation cases through Pool-Based Active Learning}

\author[1]{Jos\'e Hugo Gaspar Elsas}
\author[1,2]{Nicholas A. G. Casaprima}
\author[1,2]{Ivan F. M. Menezes}

\affil[1]{Tecgraf Institute, PUC-Rio}
\affil[2]{Department of Mechanical Engineering, PUC-Rio}

\hypersetup{
    colorlinks=true,
    linkcolor=black,
    citecolor=black,
    filecolor=black,      
    urlcolor=blue,
}

\begin{document}

\maketitle

\begin{abstract}

   A common workflow for many engineering design problems requires the evaluation of the design system to be investigated under a range of conditions. These conditions usually involve a combination of several parameters. To perform a complete evaluation of a single candidate configuration, it may be necessary to perform hundreds to thousands of simulations.  This can be computationally very expensive, particularly if several configurations need to be evaluated, as in the case of the mathematical optimization of a design problem. Although the simulations are extremely complex, generally, there is a high degree of redundancy in them, as many of the cases vary only slightly from one another. This redundancy can be exploited by omitting some simulations that are uninformative, thereby reducing the number of simulations required to obtain a reasonable approximation of the complete system. The decision of which simulations are useful is made through the use of machine learning techniques, which allow us to estimate the results of ``yet-to-be-performed" simulations from the ones that are already performed. In this study, we present the results of one such technique, namely active learning, to provide an approximate result of an entire offshore riser design simulation portfolio from a subset that is 80\% smaller than the original one. These results are expected to facilitate a significant speed-up in the offshore riser design.
      
\end{abstract}

\section{Introduction }\label{sec:Intro}

  \hspace{1em} {\color{black} Modern engineering designs place a significant emphasis in simulation techniques in order to estimate the behavior of a given system under a wide variety of circumstances. Simulating all the relevant conditions the system would be subject to, prior to experiments or deployment, is the ultimate goal.} Advances in computing power have allowed the routine use of finite element and finite volume simulations for a range of applications, even with modest computational resources. Nevertheless, it can still be challenging to run multiple simulations on a limited time horizon.
  
  Many engineering design problems require the simulation of the system under a range of boundary conditions, referred in some contexts as loading conditions, which can depend on more than one parameter. For example, to explore the design of a Y-junction mixer, {\color{black} we may need to investigate} the junction behavior as a function of flow rates in both inlets. {\color{black} This requires simulations to be performed for combinations} of parameters covering the region of interest for the two flow rates; for each combination, {\color{black} a computational fluid dynamics (CFD) calculation needs to be performed, which in itself is extremely expensive}.  {\color{black} This implies that a full exploration of both parameters can be seriously challenging, if not prohibitively expensive. } 
  
  In recent years there has been increasing interest in combining machine learning techniques with traditional engineering techniques {\color{black} , such as } physical simulation and mathematical optimization. One of the focus point for this nexus of techniques has been {\color{black} the} exploration of large parameters spaces in search for rare or counter-intuitive optimal configurations. Reviews for applications of machine learning in engineering design\cite{10.1115/1.2429697,10.1115/1.4044690} and material design \cite{LIU2017159} exemplify .
  
 {\color{black} With regard to using machine learning to explore parameter spaces in a computationally inexpensive manner}, two notable examples {\color{black} can be cited.} Peterson et al. \cite{doi:10.1063/1.4977912} used  Random Forest regression trained over a set of 60,000 simulations sampled with Latin hypercube. {\color{black} They created} a model capable of exploring the parameter space of inertial confinement fusion targets {\color{black}and found optimal, albeit } counter-intuitive configurations through optimization. Faber et al. \cite{PhysRevLett.117.135502} used ridge kernel regression to compute the formation energies of 2 million crystals from density functional theory simulations, starting from a subset of 10,000 simulations. {\color{black} However, this in itself would be extremely challenging if the simulations are required for every material.}
  
  For the design of off-shore oil risers in ultradeep-water conditions investigated in this study, most of the mechanical loading is defined by {\color{black} (i)} the sea {\color{black} currents} in the installation region of the riser, {\color{black} (ii)} the sea waves that impact the floating unit to which the riser is attached. Sea currents may change throughout the year, and {\color{black} hence, several currents must therefore be investigated}. {\color{black} There is no simple way to specifying all possible currents a riser will be subject to, therefore using a list of historically observed currents is necessary. Similarly, it is necessary to specify a list of waves using historical and meteorological data.}
  
  For this {\color{black} scenario}, the loading cases are constructed as lists of combinations of sea currents and waves, which are defined by several parameters, such as speed and direction of water at several depths, for currents, and amplitude, {\color{black} direction, and frequency in the case of waves}. Many of these combinations {\color{black} may be redundant because the mechanical tensions in the riser (the output from the analyses) of many of these simulations can be estimated from the results of other simulations}. Therefore, we seek to investigate a more efficient way to search the parameter space {\color{black} by avoiding the execution of redundant simulations}. The challenge {\color{black} here} lies in finding the most informative cases, which allow for {\color{black} a} rapid convergence of the prediction of the remainder of the list.
  
  A subject of considerable interest in this field is how to enable the mathematical optimization of the riser, while investigating the widest range of loading conditions. {\color{black} This would require executing all simulations for each configuration generated by the optimization procedure}. {\color{black} Currently, very often, this is prohibitively expensive,} owing to the very {\color{black} high} execution time required to run all the simulations. One compromise is to enable the use of optimization is to handpick the loading cases to be run during the optimization. If a critical loading case is not added to the optimization, the optimized design may fail during a later validation phase, and this would require restarting the entire optimization process. 
  
  A viable alternative to run all simulations is to dynamically choose the most informative simulations, {\color{black} and infer the results of the cases that are not executed}. This combination would allow us to indirectly use the full portfolio without the associated cost, thereby circumventing the limitations presented by manually choosing a smaller portfolio. {\color{black} Active learning is the technique we demonstrate here to perform such an inference and choice process} \cite{settles.tr09}. {\color{black} It} is a sub-field of machine learning, in which the learning algorithm supports or dictates the sampling decisions for further data, dealing with both labeled and unlabeled data.
   
  In the parlance of active learning, the simulation portfolio represents a pool of ``unlabeled" data, in which the simulations results have yet to be calculated, which must be ``queried." Each query represents an expensive operation,{\color{black} i.e., running a } simulation,  which we want to use as sparingly as possible in order to save costs. {\color{black} To achieve  the same, } we need to find an intelligent way to sample the ``yet-to-be-run" simulations in the pool. We also wish to demonstrate that the improvements are not merely a result of having more points available to estimate the un-run simulations, thereby showing that a good query criterion improves the convergence of the regression.

  {\color{black} The rest of this paper } is structured as follows: In Section \ref{sec:ActiveLearning}, we present the basic theory of active learning, including the concepts of pool-based learning, Gaussian-process regression and uncertainty Sampling. In Section \ref{sec:Methodology}, we present the offshore Riser design problem, including the most common riser configuration and the definitions of currents and waves. We conclude this section by discussing how to prepare the data to be used in the regression. In Section \ref{sec:Results}, we present the results for the active learning selection for each of the variables of interest separately and for all of them combined. Finally, in Section \ref{sec:Conclusion}, we present a perspective of the work and some closing remarks.

\section{Active Learning} \label{sec:ActiveLearning}
    
    \hspace{1em} The classical problem of active learning is designed to solve with is how to perform data acquisition in which the labeling process might be expensive, slow or complex. A comprehensive literature survey can be found in Settles \cite{settles.tr09}, on which this work is based. In the active learning problem there are four main components{\color{black}, namely (i)} the labeled training  set, {\color{black}(ii)} the machine learning model, {\color{black} (iii)} the unlabeled pool/unexplored space and {\color{black}(iv)} the ``oracle".
    
    The machine learning model performs some form of supervised learning over an initially labeled dataset, which is the training set. The trained model is used to predict the results of some desired variable,{\color{black} i.e.,} the label(s), over the unlabeled pool which can then be searched to produce a query candidate to be sent to the oracle. The oracle returns the exact label label of the queried point, which can be then added to the labeled set, {\color{black} and therefore,} used to augment the traning data. {\color{black} This cycle repeats } itself.
    
\begin{figure}[H]
        \begin{center}
          \includegraphics[width=0.75\linewidth]{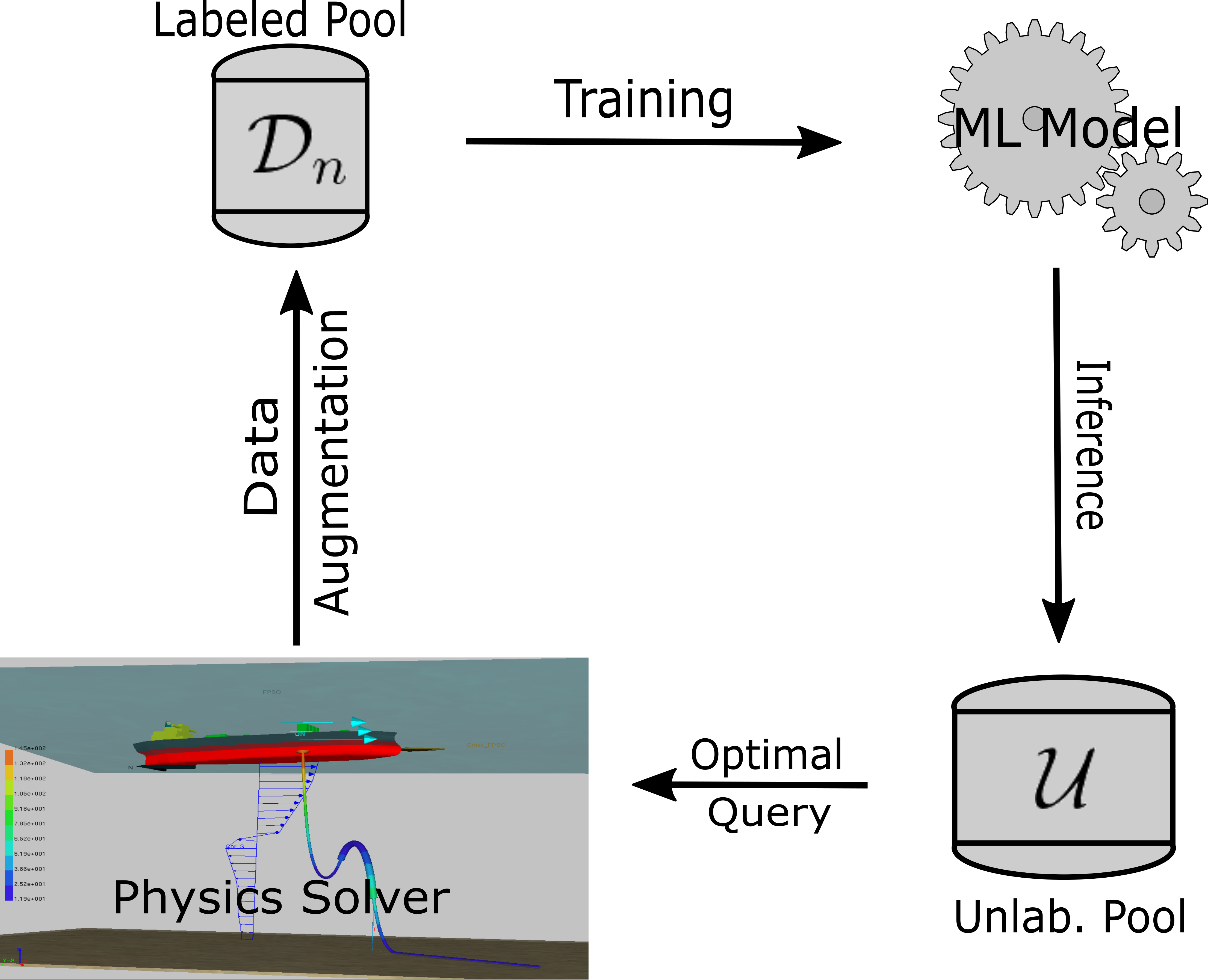}
        \end{center}
        \caption{ {\color{black} Schematic} of the active learning loop. The components of the active learning are the 'labeled pool', or dataset, of size $n$ ($\mathcal{D}_n$), 'Unlabeled pool'  ($\mathcal{U}$), the machine learning model and the Oracle. The physics solver fulfills the role of 'Oracle' in the active learning literature. Picture of the physics solver is from the program Anflex \cite{BLCP:CN026504553} for analysis of rigid and flexible risers}\label{fig:activeLearning}
      \end{figure}

    In the context of engineering design, the points represent parameters in some variable space that determine either the design variables or, in our case, the boundary conditions, which the system is subject to. The role of the oracle is performed by the simulation software, in which the boundary conditions are input. {\color{black} The software, in turn, returns} the resulting mechanical loads. The labels themselves are the results of the simulation, which {\color{black} in the case of our problem are not labels} in the sense of classification problems, but rather target variable values as in regression problems. Finally, the un-labelled pool is the list of yet-to-be-run simulations. A complete diagram of the active learning loop can be seen in Figure \ref{fig:activeLearning}.
    
    The initial training set can be formed by randomly selecting cases from the complete list or by running simulations sampled by some {\color{black} type of a} ``design of experiment", {\color{black} for example, the } Latin Hypercube sampling \cite{McKay1979}. This cycle allows us to sequentially run the simulations {\color{black} in such a way that the quality of the predictions of the yet-to-be-run simulations (unlabeled pool) is maximized.} To be able to predict the value of the labels over the un-labeled pool, we need a machine learning model to perform the regression over those points, which we apply Gaussian process regression.
    
  \subsection{Gaussian Processes Regression}\label{subsec:GP}
  
    \hspace{1em}  Gaussian Process regression \cite{Rasmussen:2005:GPM:1162254} is an efficient technique for interpolations that approach arbitrary functions, while providing estimations for both the function and the uncertainty in such estimation. In the context of active learning, the ability to provide easy estimations of uncertainty is particularly useful, as will be discussed in Section \ref{subsec:UnSampling}. In {\color{black} this study}, Gaussian process regression performs the role of the machine learning model in the active learning procedure. 
    
    The Gaussian process regression expresses the problem of estimating the function value by assuming that the objective function is drawn from a Gaussian process. This means that the distribution over any set of values of the function on $n$ points $\mathcal{D}_n = \{f({\bf x}_1),f({\bf x}_2),...,f({\bf x}_n)\}$, is a joint multivariate normal distribution, $\mathcal{N}(\mu_n,\Sigma_n)$. {\color{black} The} mean vector and covariance matrix defined as $(\mu_n)_k = m({\bf x}_k) = m({\bf x})_k$ and $(\Sigma_n)_{ij} = k({\bf X},{\bf X})_{ij} = k({\bf x}_i,{\bf x}_j)$, respectively. The function $m({\bf x})$ is the mean, and $k({\bf x},{\bf x}')$ the covariance of the Gaussian process. The statement that $f({\bf x})$ is modeled by the Gaussian Process is written as $f({\bf x}) \sim \mathcal{GP}(m({\bf x}),k({\bf x},{\bf x}'))$.
      
    The Gaussian process for regression is performed by computing the conditional probability, $P(f({\bf x}) | \mathcal{D}_n)$, of the value of a point $f({\bf x})$ given the value of the previous $n$ points, $\mathcal{D}_n$. The conditional probability is itself a normal distribution; its parameters are given by
        
    \begin{eqnarray}
        f({\bf x}) | \mathcal{D}_n &\sim&  \mathcal{N}(\mu_n^*({\bf x}),\sigma_n^*({\bf x})) \label{eq:GPreg}\\
          \mu_n^*({\bf x}) &=& m({\bf x}) + k({\bf x},{\bf X})k({\bf X},{\bf X})^{-1} (Y - m({\bf X})) \label{eq:mu_n}\\
           \sigma_n^*({\bf x}) &=& k({\bf x},{\bf x}) - k({\bf x},{\bf X})k({\bf X},{\bf X})^{-1} k({\bf X},{\bf x}) \label{eq:sigma_n}
    \end{eqnarray}       
    where $Y$ is the vector of observed evaluations, $Y_k = f({\bf x}_k)$, or $Y_k = f({\bf x}_k) + \epsilon_k$ in the case of noisy evaluations with iid noise, $\epsilon_k \sim \mathcal{N}(0,\sigma^2)$; $k({\bf X},{\bf x})$ is a column vector in $k({\bf X},{\bf x})_k = k({\bf x}_k,{\bf x})$; $k({\bf x},{\bf X})$ is the transpose vector; given the covariance $k$ used to model the objective function, $k({\bf X},{\bf X}) = \Sigma_n$ is the covariance matrix associated with dataset $\mathcal{D}_n$. The quality of the interpolation practically depends on the choice between {\color{black} the} functions $m$ and $k$ in encoding the assumption on the behavior, especially the smoothness of the modeled function, $f({\bf x})$.
      
    The covariance alone {\color{black}, is very often}, sufficiently powerful to model the entire function; hence, a common choice is to set $m({\bf x}) = 0$. Nonetheless, the choice of covariance function $k({\bf x},{\bf x}')$ cannot be avoided and constitutes one of the modeling building blocks for the Gaussian process regression, and {\color{black} consequently}, the success of the active learning process.
    
    For Gaussian Process Regression, training {\color{black} approximately} corresponds to computing the {\color{black} product} $k({\bf x},{\bf X})k({\bf X},{\bf X})^{-1}$. For other machine learning models this could correspond to the gradient descent procedure, as is the case of neural networks, or any other training procedure. The key idea is to train over the set of points that have been already labeled, i.e. the points over which the simulation have already been run.
    
    Gaussian Process regression is particularly effective in cases {\color{black} where the number of points to be used in the training set is small or limited}, under which most learning algorithms struggle. Because the goal of accelerating engineering design by selecting most informative simulations entails having the fewest simulation available for regression, it is reasonable to use the most efficient method available.  One of the main limitations of Gaussian Process regression, {\color{black} owing to its being an} intrinsically non-parametric learning method, is the sharp rise in computational cost as the number of training points increases. {\color{black} However, this} is not a problem in our case. 
    
  \subsection{Uncertainty Sampling}\label{subsec:UnSampling}
  
    \hspace{1em}  Once the model is trained, it can be used to predict the values of the results of the physics simulations, i.e. to perform inference of the labels, over the unlabeled pool of possible yet-to-be-run cases. Each inference produces a prediction of the value of the physical variable computed by the physics solver {\color{black} in addition to} an implied uncertainty. This uncertainty express, from a Bayesian point of view, a confidence interval for the predicted variables. {\color{black} This} entails the amount of information that can be gained by running the respective simulation.
    
    To produce a candidate query, one must choose at least one candidate from the unlabeled set $\mathcal{U}$, which essentially amounts to {\color{black} optimizing} some selection criterion over the set $\mathcal{U}$.
    
    A very simple and commonly used query criterion is called ``uncertainty sampling'' \cite{10.5555/188490.188495}, in which the inferred uncertainty by the machine learning model is used as the objective function to optimize, which is a form of Information-Based Objective Function, {\color{black} as described in the seminal work} by MacKay\cite{macKay92}. Therefore, for our finite dataset $\mathcal{U}$, the optimal query amounts to finding the point to be queried with the maximum inferred uncertainty ${\bf x}^*_{\rm us}$:
    
    \begin{equation}
       {\bf x}^*_{\rm us} = {\rm argmax}_{{\bf x}_i \in \mathcal{U}}\ \sigma_n({\bf x}_i)
    \end{equation}
    
    For classification problems, alternative objective functions may include classification entropy \cite{shannonEntropy} which is a measure of information, to encode a given distribution from the data already available.  {\color{black} If a significant amount of information is not required to encode a new case from the old ones, the new case may not be very informative if it were to be acquired. In this study, we only used uncertainty sampling, which is elaborated in Section \ref{sec:Results}.}
    
\section{Methodology}\label{sec:Methodology}
  
  \hspace{1em} To demonstrate the effectiveness of the loading case selection we ran the procedure on a reference dataset, in which all simulations were run and the active learning loop was executed. The difference between predicted and exact values {\color{black} in} the unlabeled set $\mathcal{U}$ was computed at each iteration. With the exact value{\color{black}s}, it is possible to compute how the implied uncertainty, represented by the predicted standard deviation, relates to the error of the machine learning prediction. If it {\color{black} were} possible to bound the error by a function of the implied uncertainty, we can deduce that it is possible to omit simulation as long as the error is kept under control.
  
  In Section \ref{subsec:SLWR} we will discuss the physical system used as an example {\color{black} in this study}, which is a Steel Lazy Wave Riser (SLWR) in a ultra-deep water environment, including references for the definitions of the physical variables of interest and the physics solver used to produce the simulation data. In Section \ref{subsec:LoadingCases}, we will present details on how the input parameters for the simulations {\color{black} were} defined, including how sea currents and waves are parametrized.
  
  {\color{black} Next, it is essential to format the data to be used in the machine learning in such a way that the feature vectors ${\bf x}$ and target labels ${\bf y}$ can be fed to the model}. The feature vectors' ${\bf x}$ component are, mostly, normalized input parameters, and we present the normalization used in this work in Section \ref{subsec:dataFormating}. Targets are also normalized for some of the variables, details will be discussed in Section \ref{sec:Results}.
  
  \subsection{Steel Lazy Wave Risers}\label{subsec:SLWR}
    
\begin{figure}[H]
        \begin{center}
          \includegraphics[width=0.8\linewidth]{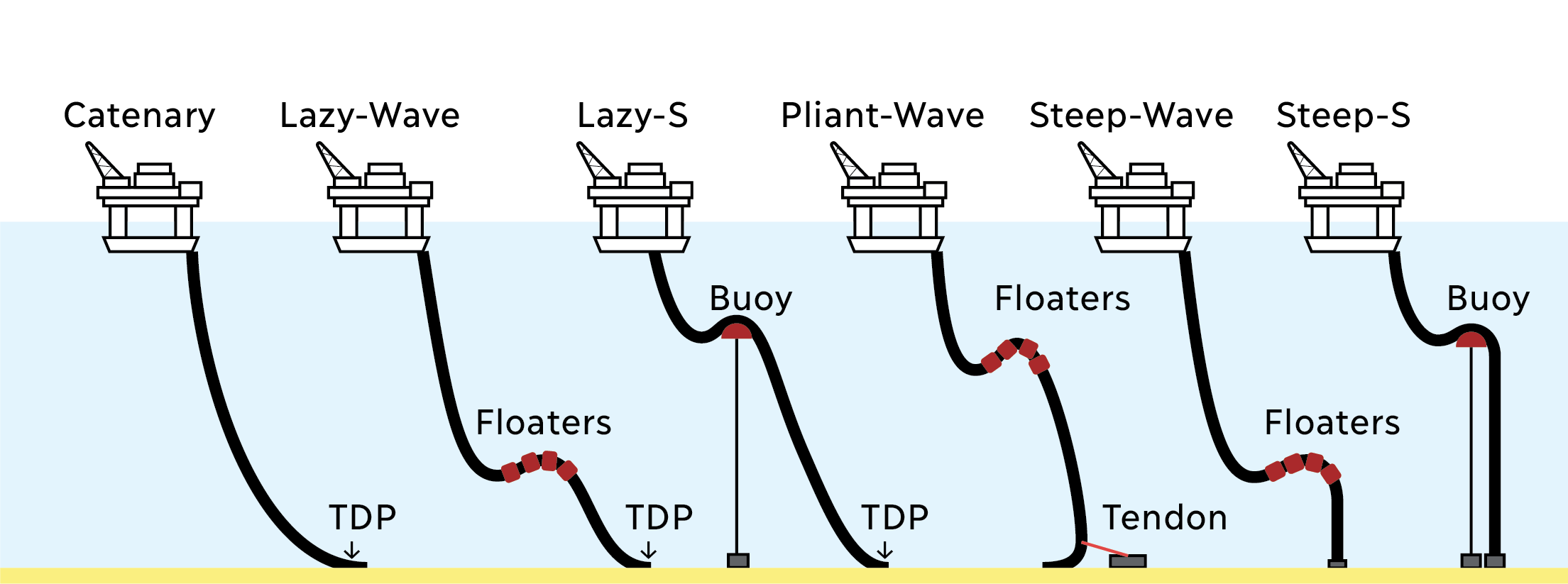}
        \end{center}
        \caption{ Riser configurations, taken from Cardoso et al.\cite{cardoso2019optimization} adapted from Clausen and D'Souza\cite{clausen2001}.}\label{fig:riserconfig}
      \end{figure}

    \hspace{1em} In this study, we apply active learning to the design of production riser in a SLWR configuration. Production risers are pipes, which can be rigid or flexible, used to flow oil and gas from the wellhead to the floating unit. There are multiple possible deployable configurations (Figure \ref{fig:riserconfig}), and the decision {\color{black} as to which one fits best} is based on water depth, environmental conditions and determined by waves and currents, {\color{black} as well as cost}. 

    The simplest and easiest to deploy is the free hanging catenary, in which the risers are simply attached to the floating unit {\color{black} at} one end and to the wellhead {\color{black} at} the opposite end without any buoys. As the water depth increases and the environmental conditions get harsher, this configuration may no longer be feasible. {\color{black} This is because} the mechanical stresses increase {\color{black} and} lead to structural problems, such as buckling in the region where the riser touches the seabed, known as the touchdown zone (TDZ). Studies indicate \cite{jacob1999alternative} that the riser configuration that include buoyancy elements, {\color{black} e.g.,} SLWR, have milder {\color{black} stresses} than those without buoyancy, {\color{black}e.g.,} free hanging catenary. Therefore, configurations such as the SLWR are more suitable for deployment in conditions found in ultra-deep waters.

\begin{figure}[H]
        \begin{center}
          \includegraphics[width=0.8\linewidth]{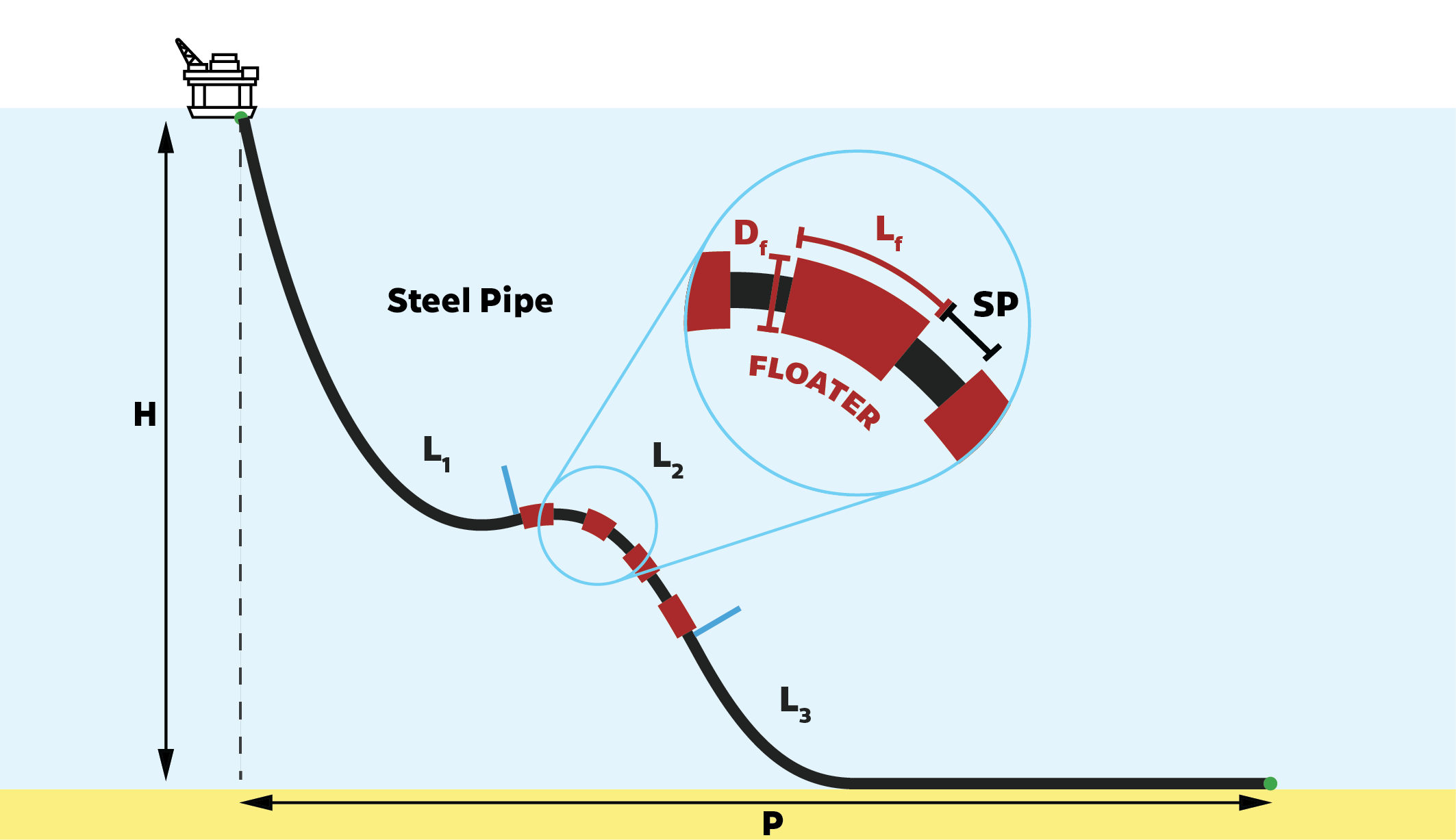}
        \end{center}
        \caption{ SLWR configuration }\label{fig:slwr}
      \end{figure}

    The SLWR configuration (Figure \ref{fig:slwr}) is defined mainly by the top angle ($\theta_{\rm top}$), the length of the top segment ($L1$), the middle segment where the buoys are located ($L2$), the bottom length ($L3$) and the buoys' dimensions, such as their length ($L_f$), diameter ($D_f$) and  inter-buoy spacing ($SP$). In order for a configuration to be considered feasible, it must keep key mechanical stress metrics under control in a variety of different environmental conditions.
 
   To assess the behavior of the riser under these conditions, in principle, simulations must be run for each of the possible combination of waves and currents that are on record. The currents and waves  investigated in the riser design are specified by historical data and are not controllable by the engineers responsible for the project, which motivates the application of pool-based active learning to this problem. The number of combinations of waves and currents can exceed 1000 finite element analysis.
 
   {\color{black} Specifically, } the case presented on this paper is a SLWR deployed in a location with a water depth of 2200 m and the horizontal distance between the connections on the floating unit to the connection on the wellhead is 3001 m. The riser studied is composed {\color{black} of} rigid pipes with properties described on table \ref{tab:pipeprop}:
 
         \begin{table}[H]
         \begin{center}
         \begin{tabular}{|c|c|} \hline 
         Inner diameter    & $0.35560\ m$           \\ \hline
         Outer diameter    & $0.43560\ m$           \\ \hline
         Coating Thickness & $0.048\ m$             \\ \hline
         Young's modulus   & $207\ {\rm GPa}$       \\ \hline
         Density           & $7919\ kg/m^3$			\\ \hline
         Yield strength    & $404.16\ {\rm MPa}$    \\ \hline
         Tensile strength  & $483.84\ {\rm MPa}$    \\ \hline
         \end{tabular}
         \end{center}
        
        \caption{Pipe properties} \label{tab:pipeprop}
        \end{table}
  
  \subsection{Configurations}\label{subsec:configurations}
     \hspace{1em} Each unlabeled pool is generated from a riser configuration by performing a series of simulation with varying loading conditions. For the same loading condition, different riser configurations perform differently, which therefore allows us to explore how dependency on the particular configuration affects performance of the active learning loop. Using the parameters shown in Figure \ref{fig:slwr}, the Risers we used to generate the data are described in Table \ref{tab:riser_configs}:
     
         \begin{table}[H]
         \begin{center}
         \begin{tabular}{|c|c|c|c|c|c|c|} \hline 
            Param. & Conf. 0 & Conf. 1 & Conf. 2 & Conf. 3 & Conf. 4 & Conf. 5 \\ \hline
            $\theta_{\rm top} ({}^o)$ & 7.0 &  7.919 &  7.923 &  7.997 &  7.989 & 8.000 \\ \hline
            $L_3 ({\rm m})$ & 1421.500 & 1019.069 &  973.240 &  676.622 &  855.062 &  108.000 \\ \hline
            $L_2 ({\rm m})$ & 828.000 &  463.364 &  502.765 &  541.291 &  563.301 &  991.379 \\ \hline
            $L_1 ({\rm m})$ & 2325.000 & 2238.995 & 2265.995 & 2252.889 & 2347.458 & 2480.278 \\ \hline
            $D_f ({\rm m})$ & 2.400 &  2.788 & 2.808 & 2.795 & 2.838 & 3.000 \\ \hline
            $L_f ({\rm m})$ & 3.000 &  2.585 & 2.564 & 2.437 & 2.538 & 2.293 \\ \hline
            $SP ({\rm m})$ & 12.000 & 12.897 & 12.925 & 13.300 & 13.439 & 14.647 \\ \hline
         \end{tabular}
         \end{center}
        
         \caption{Parameters characterizing the Riser Configurations utilized.} \label{tab:riser_configs}
         \end{table}

  \subsection{Loading Cases}\label{subsec:LoadingCases}
     \hspace{1em} Loading cases are used to establish the environmental impacts on the risers and are therefore, vital to the accuracy of the simulations. A loading case is a combination of {\color{black} a current and a wave}, which defines one of the states to be studied. {\color{black} For a study to be} representative of a real deployed case, it should contain information from the location where the riser is to be deployed, and the results, such as mechanical stresses and norm factors, should all be under a safety criteria or material limits.

    \subsubsection{Currents}\label{subsubsec:Currents}
\begin{figure}[H]
        \begin{center}
          \includegraphics[width=0.4\linewidth]{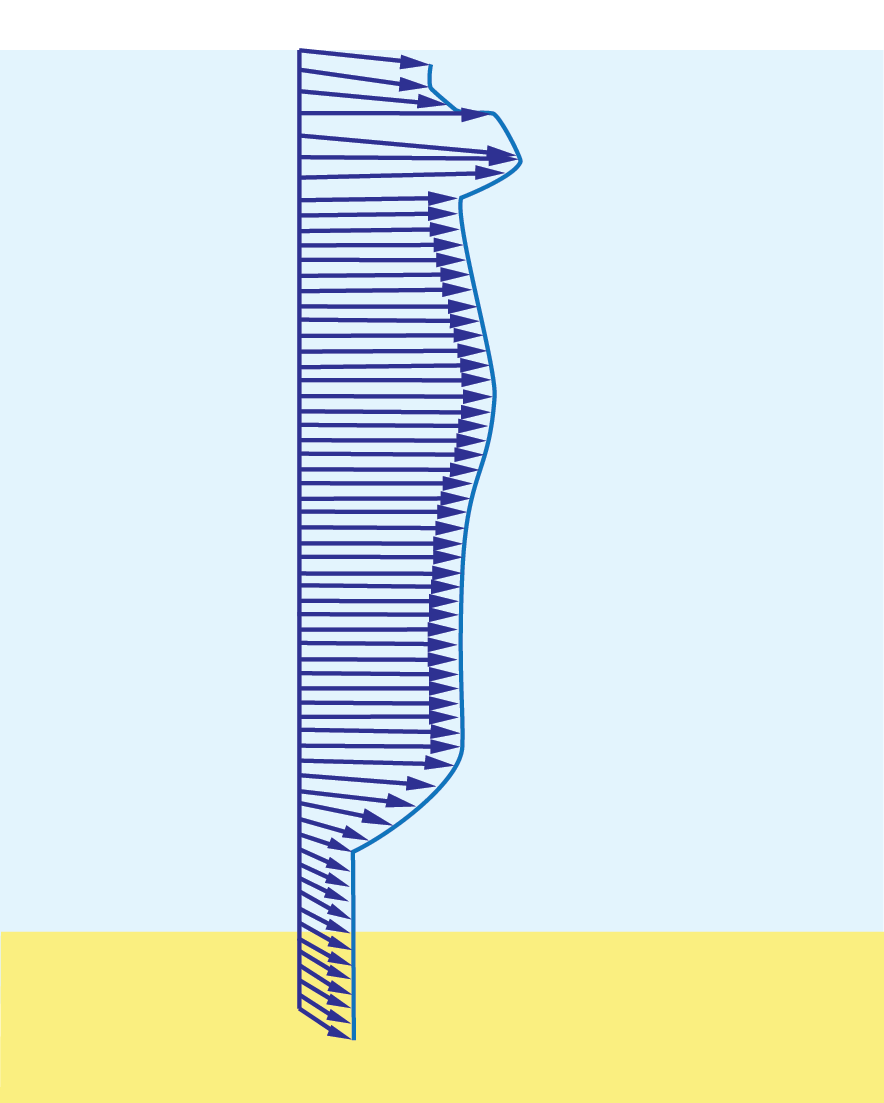}
        \end{center}
        \caption{ Illustrative representation of a current velocity profile  }\label{fig:currentvp}
      \end{figure}

      \hspace{1em} Currents are defined by the velocity profile over the water depth, as {\color{black} shown } in Figure \ref{fig:currentvp}. The velocity profile is expressed discretely and three main attributes must be defined, {\color{black} namely} the water depth of each step of the discretization, and the corresponding direction and magnitude of the water velocity. An example of a sea current is shown in Figure \ref{fig:currentvp}, and the actual data defining a current is given in Table 2. For the purpose of this work, currents are considered as static loads, which do not vary during the simulation.

        \begin{table}[H]
        \begin{center}
        \begin{tabular}{|c|c|c|c|} \hline 
        Depth (m) & Velocity (m/s) & Direction & Angle ($\theta$) \\ \hline
        0         & 0.46           & SW        &     $225^{o}$    \\ \hline
        50        & 0.46           & SW        &     $225^{o}$    \\ \hline
        100       & 0.46           & SW        &     $225^{o}$    \\ \hline
        150       & 0.42           & SW        &     $225^{o}$    \\ \hline
        200       & 0.40           & SW        &     $225^{o}$    \\ \hline
        250       & 0.39           & WSW       &     $247.5^{o}$  \\ \hline
        300       & 0.39           & WSW       &     $247.5^{o}$  \\ \hline
        350       & 0.37           & W         &     $270^{o}$    \\ \hline
        375       & 0.36           & W         &     $270^{o}$    \\ \hline
        800       & 0.41           & NW        &     $315^{o}$    \\ \hline
        1200      & 0.32           & NW        &     $315^{o}$    \\ \hline
        1600      & 0.20           & NW        &     $315^{o}$    \\ \hline
        2000      & 0.20           & N         &     $0^{o}$      \\ \hline
        2200      & 0.20           & N         &     $0^{o}$      \\ \hline
        \end{tabular}
        \end{center}
        
        \caption{Example table defining a current} \label{tab:exCurrent}
        \end{table}
        
    \subsubsection{Waves}\label{subsubsec:Waves}
    
	   \hspace{1em} Oceanic waves on the surface are usually irregular, varying in length, direction, and height. To model such behavior, waves are treated as stochastic process with a parametrized spectrum. One of the most frequently used  spectra for modeling is the Joint North Sea Wave Observation Project (JONSWAP) model \cite{jonswap73}. The power spectrum of the wave in this model is defined as: 
	
        \begin{eqnarray}
          S(\omega) &=& \frac{\alpha g^2}{\omega^5} \exp\left[-\frac{5}{4} \left(\frac{\omega_p}{\omega}\right)^4\right] \gamma^r \\
          r &=& \exp\left[-\frac{(\omega-\omega_p)^2}{2\sigma^2 \omega^2_p}\right]
        \end{eqnarray}
    
      This model has free parameters given by a base coefficient $\alpha$, dominant angular frequency $\omega_p$, secondary base $\gamma$ and secondary width $\sigma$, along the direction $\theta$ from which the wave is traveling from. $\gamma$ is sometimes called the enhancement factor.
      
      In this study, the waves were defined by inserting five main properties: the wave's height, period, direction and spectrum. With this information, the simulation algorithm is capable of calculating the spectrum coefficients. An example of a set of wave parameters is shown in table \ref{tab:exWave}:
      
        \begin{table}[H]
        \begin{center}
        \begin{tabular}{|c|c|c|c|c|} \hline 
        Height (m) & Period (s) & Azimuth ($\theta_{\rm az}$) & $\alpha$ & $\gamma$ \\ \hline
		1.70       & 4          & 180${}^o$     & 0.037856 & 3.240175 \\ \hline
        \end{tabular}
        \end{center}
        
        \caption{Example table defining a wave, which are the height $H$, dominant period $T_p$, azimuthal angle $\theta$, and power spectrum parameters $\alpha$ and $\gamma$}  \label{tab:exWave}
        \end{table} 
        
   \subsubsection{Target Variables}
   
     \hspace{1em} The result of the simulation of the riser dynamics are two sets of variables related to the mechanical tension of the riser during the simulation. The first variable is the maximum Axial tension (${\rm FX}$) on the top of the riser  over the course of the simulation. This variable is meant to be negative denoting the fact that the riser should always pulls down on the floating unit. 
     
     The second variable returned by the simulation {\color{black} was} the maximum normalized tension, named DNVUF-201, along the entire riser, computed over the course of the simulation. The pipe elements are subjected to bending moment and effective tension; these {\color{black} were}  used to define a variable called ``DNV Utilization Factor 201" (DNVUF-201), which a combined loading criteria defined by the offshore standard DNVGL-ST-F201 \cite{DNV-OF-F201} as follows:
     
          \begin{equation}
             \begin{dcases*} 
              \gamma_{SC} \gamma_m \left[\frac{|M_d|}{M_k} \sqrt{1-\left(\frac{p_i-p_0}{p_b}\right)^2} + \left(\frac{T_e}{T_k}\right)^2\right] + \left(\frac{p_i-p_o}{p_b}\right)^2 {\rm if}\ p_i > p_o\\ 
              (\gamma_{SC}\gamma_m)^2 \left\{ \left[\frac{|M_d|}{M_k} + \left(\frac{T_e}{T_k}\right)^2\right]^2 + \left(\frac{p_o-p_{\rm min}}{p_c}\right)^2\right\}{\rm if}\ p_i \leq p_o
            \end{dcases*} 
          \end{equation}
        
        where $\gamma_{SC} =$ material safety class factor, $\gamma_m =$ material resistance factor, $M_d = $ bending moment, $M_k = $ plastic bending moment, $T_e = $ effective tension, $T_k = $ plastic axial force, $p_i = $ internal pressure, $p_o = $ external pressure, $p_b = $ burst pressure (for $p_i > p_o$), $p_c = $ collapse pressure (for $p_i < p_o$) and $p_{\rm min} = $ minimum internal pressure. In a proper installation, in which no excessive tension is applied to the riser, it is expected that the maximum of DNVUF-201 satisfies $\max$ DNVUF-201 $\leq 1$ over the entire riser length during the entire simulation, for all different loading cases. 
  
  \subsection{Data-Formatting}\label{subsec:dataFormating}
  
    \hspace{1em} To feed the machine learning model, the data must be in an appropriate format for use in Gaussian process regression. To compute the covariance values using a standard kernel, such as Gaussian or Mat\`ern kernels, categorical features must be converted into some form of real variable value. In our case, the major transformation was to convert compass direction into {\color{black} angular} information, and then into horizontal and vertical components of a vector. 
    
    The target variables for these simulations are the values of the DNVUF-201 and the value of {\color{black} FX axial force} at the top section of the riser, where the pipe connects to the floating unit. To this end, we normalized the FX variables over the dataset $\mathcal{D}_n$, meaning that we repeated the normalization at each iteration of the active learning process. {\color{black} In} doing so we only used the data of the simulations that would have been performed for normalization. 
    
    We did not normalize the DNVUF-201 for the end case. Normalizing the data either through computing its logarithm or through typical mean subtraction and standard deviation normalization did not improve the performance of the active learning algorithm and {\color{black} hence,} was not considered. 
    
    \subsubsection{Currents and Waves}\label{subsubsec:dataCurrents}
      
      \hspace{1em} As mentioned previously, currents are defined by specifying {\color{black} their} velocity field at different depths. In its original form, the data format {\color{black} cannot be easily processed} by a machine learning model. To address this shortcoming, we transformed the information in the table into a single array that {\color{black} contained} the information from all three columns. For vector fields, it is natural to use the $L_2$-norm to compute the similarity of two  fields. {\color{black} A} sea current is a vector field ${\bf u}: [a,b] \rightarrow \mathbb{R}^2$, which defines the x and y components of the water velocity at each depth. 
      
      For a velocity field that is specified on a set of nodes $\{x_0,x_1,...,x_n\} \subset [a,b]$, in which ${\bf u}_i = {\bf u}(x_i)$, it is possible to approximate the $L_2$-norm of the velocity vector ${\bf u}$ by a Riemann sum as:
      
      \begin{eqnarray}
         & & \int_a^b ||{\bf u}(x)||^2 dx  \approx  \sum_{i=0}^{n-1} \frac{1}{2} (x_{i+1}-x_i) (||{\bf u}_i^2|| + ||{\bf u}_{i+1}^2||) = \label{eq:velRescale} \\
        &=& \left|\left|\sqrt{\frac{x_1-x_0}{2}} {\bf u}_0\right|\right|^2 + \sum_{i=1}^{n-1} \left|\left|\sqrt{\frac{x_{i+1}-x_{i-1}}{2}}{\bf u}_i \right|\right|^2 + \left|\left|\sqrt{\frac{x_n-x_{n-1}}{2}} {\bf u}_n\right|\right|^2  \nonumber \\
        &=&  \sum_{i=0}^{n} ||{\bf v}_i||^2 \nonumber 
      \end{eqnarray}
      
      Equation \eqref{eq:velRescale} presents a natural way to re-write the information contained in tables such as in Table \ref{tab:exCurrent}. Compass points are converted into angles $\theta$, and then horizontal and vertical components of the velocity field are computed as $u_x = u \cos{\theta}$ and $u_y = u \sin{\theta}$, respectively, where $u$ in the absolute value of the velocity, as informed by the column ``velocity". Then, it is necessary to re-scale the velocity components according to the coefficients in Equation \eqref{eq:velRescale}. This allow to define x and y values for velocity components at different depths, i.e. $\{{\rm vx}0, {\rm vx}50, {\rm vx}100, ..., {\rm vx}2200\}$ and $\{{\rm vy}0, {\rm vy}50, {\rm vy}100, ..., {\rm vy}2200\}$. {As in our case}, all currents are specified at the same depths, the information about the depth itself can be made purely positional, thereby allowing the first column to be suppressed after the re-scaling has been performed.
      
      The set $\{{\rm vx}0, {\rm vx}50, ..., {\rm vx}2200, {\rm vy}0, {\rm vy}50, ..., {\rm vy}2200\}$ comprises the sea current contribution to the feature vector utilized in the training and inference steps for each of the loading cases. The values obtained for the current feature vectors were further normalized to make the means zero and standard deviations unitary over the set of loading conditions.
      
      As mentioned in Section \ref{subsubsec:Waves}, the waves {\color{black} were} defined by the JONSWAP model parameters. An example wave can be seen in Table \ref{tab:exCurrent}. The {\color{black} dominant} angular frequency {\color{black} is} computed as $\omega_p = 2 \pi / T_p$. To prepare the data for the machine learning model, we absorbed the direction information by multiplying by coefficient $\alpha$ as follows: $\alpha_x = \alpha \cos \theta_{\rm az}$ and $\alpha_y = \alpha \sin \theta_{\rm az}$. The {\color{black} remaining parameters} are used {\color{black} without any modification}. Therefore, the feature vector components related to the sea wave {\color{black} were} $\{\alpha_x,\alpha_y,\sigma,\gamma,\omega_p\}$.

  \subsection{Implementation details} \label{subsec:Implementation}
     
     \hspace{1em} To produce the base data {\color{black} comprising} the results of the mechanical response of the risers to the several sea current and waves cases, we utilized the Anflex simulation software \cite{BLCP:CN026504553}. Anflex is a computational program for nonlinear, both static and dynamic analysis of risers based on the Finite Element Method.
     
     The active learning loop, including the pre-processing of the simulation parameters to turn them into proper feature vectors, was {\color{black} written} using a combination of the numpy \cite{oliphant2006guide}, pandas \cite{reback2020pandas,mckinney-proc-scipy-2010} and scikit-learn \cite{scikit-learn} opensource python libraries. {\color{black} An example dataset}, together with Jupyter notebooks \cite{PER-GRA:2007} that implement the active learning procedure are available in the url: \url{https://git.tecgraf.puc-rio.br/jhelsas/active-learning-loading-case-selection}
     
     The kernel function utilized was a pure Gaussian, in which the width and the constant coefficient are available for the library to optimize over the training set. The errors {\color{black} were} computed after transforming the data back to the original normalization of the data, instead of using the normalized data. 
     
     The dataset as a whole was comprised of 526 simulations for 6 different riser configuration. The feature vector for each loading case was had 33 components, 5 coming from the definition of the wave and 28 coming from the definition of the current. There were 6 target variables that models were trained to predict, which correspond to the DNVUF-201 and ${\rm FX}$ variables for each loading condition. The training dataset increases along the active learning loop, starting at 25 elements and going up to 325 elements. The test dataset, which is the unlabeled set $\mathcal{U}$, shrinks from 501 elements to 201 elements.

\section{Results}\label{sec:Results}
   
   \hspace{1em} We divide the results in two case studies. In Section \ref{subsec:singleVar} we {\color{black} present an investigation of the case}, in which the uncertainty and selection is done for each target variable separately. {\color{black} Next}, in Section \ref{subsec:multiVar}, we {\color{black} present} the case in which one selection {\color{black} was performed} for all variables simultaneously and the uncertainty {\color{black} was} computed jointly for all variables. {\color{black} The second case was more reflective of reality, because the simulation returns more than one variable of interest.}.
   
   {\color{black} Finally}, in Section \ref{subsec:randomActive}, we {\color{black} present our investigation on} whether the convergence {\color{black} was on account of a} smart choice from the unlabeled dataset or {\color{black}was} only an volume effect {\color{black} owing} to the number of simulations run. To do so we compared the measures of error {\color{black} with those of} the case, in which the sampling {\color{black}was} purely random {\color{black} and therefore}, {\color{black}did not have} any contribution from the uncertainty sampling component.
   
   For a single variable $k$, the {\color{black}machine learning} model {\color{black}predicts} $\mu_n^k({\bf x}_i),\sigma_n^k({\bf x}_i)$  for ${\bf x}_i \in \mathcal{U}$ which have several uncertainties, one for each physical quantity returned by the simulation. To be able to perform an uncertainty sampling, a single variable must be chosen to represent the quantity {\color{black} being} optimized. In principle, all variables can be potential candidates. This poses a difficulty in extending the sampling procedure to more than one variable.
   
   {\color{black} Therefore}, instead of {\color{black}attempting some sort of} multi-objective optimization, {\color{black}we computed} the simplest variable that aggregate the uncertainty from all variables, {\color{black} i.e.,} the geometric mean of the uncertainties $\bar{\sigma}_n({\bf x}_i) = \left(\Pi_{k=0}^K \sigma_n^k({\bf x}_i) \right)^{1/K}$. This {\color{black} allowed} us to perform regular uncertainty sampling over this averaged quantity {\color{black} in} the same way we would {\color{black}for} a regular uncertainty. We demonstrate in Section \ref{subsec:multiVar} that this choice did not degrade the performance of the active learning procedure in converging the regression over the unlabeled dataset $\mathcal{U}$. 
   
   We present, in Table \ref{tab:predTable}, a sample of the values of all target variables on $\mathcal{U}$ {\color{black}mentioned in} Section \ref{subsec:multiVar}. {\color{black} Here,} both the exact and predicted values are shown side by side. It is clear from Table \ref{tab:predTable} that it is possible to get reasonably good approximations to the values of the simulations through the inference process, which is one of the central {\color{black}tenets} of this work. 
   
   \begin{table}[H]
    \begin{center}
   \begin{tabular}{|c|c|c|c|c|c|} \hline
    \multirow{2}{*}{} & \multicolumn{5}{c}{DNVUF-201 criterion Empty} \vline  \\ \hline
    Case & Exact & Predicted & Error (Abs.) & Error (\%) & Std.Dev. \normalsize \\ \hline    
    \textbf{42}  & 0.477585 & 0.496649 & 0.019063 & 3.991616 & 0.016616 \\ \hline
    \textbf{121} & 0.548855 & 0.528666 & 0.020189 & 3.678384 & 0.014957 \\ \hline
    \textbf{167} & 0.515923 & 0.539383 & 0.023460 & 4.547210 & 0.016632 \\ \hline
    \textbf{181} & 0.662742 & 0.637638 & 0.025104 & 3.787917 & 0.018406 \\ \hline
    \textbf{250} & 0.572730 & 0.592696 & 0.019966 & 3.486091 & 0.015458 \\ \hline
    \textbf{312} & 0.667698 & 0.665142 & 0.002556 & 0.382833 & 0.015781 \\ \hline
    \textbf{331} & 0.645096 & 0.644987 & 0.000109 & 0.016876 & 0.015169 \\ \hline
    \textbf{343} & 0.628863 & 0.623250 & 0.005613 & 0.892565 & 0.012740 \\ \hline
    \textbf{394} & 0.627311 & 0.620269 & 0.007043 & 1.122667 & 0.012472 \\ \hline
    \textbf{399} & 0.577041 & 0.601660 & 0.024619 & 4.266426 & 0.015971 \\ \hline
    \multirow{2}{*}{} & \multicolumn{5}{c}{FX axial tension Empty} \vline \\ \hline
    Case & Exact & Predicted & Error (Abs.) & Error (\%) & Std.Dev. \normalsize \\ \hline
    \textbf{42}  & -3743.67 & -3755.80 & 12.1212  & 0.323780 & 21.1400 \\ \hline
    \textbf{121} & -2456.17 & -2406.77 & 49.4048  & 2.011456 & 37.6659 \\ \hline
    \textbf{167} & -3179.94 & -3113.61 & 66.3339  & 2.086011 & 54.6702 \\ \hline
    \textbf{181} & -4479.17 & -4268.80 & 210.3750 & 4.696736 & 27.6913 \\ \hline
    \textbf{250} & -4080.17 & -3968.35 & 111.8178 & 2.740516 & 50.7698 \\ \hline
    \textbf{312} & -3540.84 & -3589.40 & 48.5529  & 1.371224 & 57.7834 \\ \hline
    \textbf{331} & -3838.46 & -3827.99 & 10.4802  & 0.273032 & 19.4366 \\ \hline
    \textbf{343} & -4451.07 & -4438.07 & 12.9975  & 0.292009 & 31.0689 \\ \hline
    \textbf{394} & -3156.19 & -3217.21 & 61.0162  & 1.933222 & 41.9970 \\ \hline
    \textbf{399} & -4310.14 & -4294.23 & 15.9168  & 0.369288 & 40.3006 \\ \hline
   \end{tabular}
    
    \end{center}
   
   \caption{Comparsion of exact and predicted values for the DNVUF-201 {\color{black}and} ${\rm FX}$, jointly sampled together with the other variables, as {\color{black}explained} in Section \ref{subsec:multiVar}. {\color{black} The} results are presented for an empty riser interior. The table was extracted from data, in which $\mathcal{D}_n$ {\color{black}was} initialized with 25 randomly selected cases and {\color{black} the active learning loop was then iterated for another 75 times}. {\color{black} This took the total number of points on the ``ran simulations" dataset} $\mathcal{D}_n$ to 100 points.}\label{tab:predTable}
   \end{table}
    
   To quantify the quality of the approximation and measure the speed of convergence, we {\color{black} computed} several deviation measures from the data; {\color{black} firstly, we compared} the root mean square (RMS) error against the root mean square implied standard deviation (Std.Dev.) and maximum standard deviation, in which the averaging process was {\color{black} performed} over the yet-to-be run simulations $\mathcal{U}$. If it {\color{black}was} possible to approximate or bound the RMS error, which depends on data beyond the labeled dataset $\mathcal{D}_n$, by {\color{black}an} affine combination of {\color{black} the} RMS and maximum implied Std.Dev., then controlling the desired error over $\mathcal{U}$ {\color{black}was} possible with only the data on the labeled dataset $\mathcal{D}_n$. To fully control the error, it {\color{black} was} also necessary to bound the maximum error over $\mathcal{U}$, using one such affine combination. Sections \ref{subsec:singleVar} and \ref{subsec:multiVar} establish exactly these bounds for the cases, in {\color{black}which} each variable {\color{black}was} queried separately, {\color{black} whereas, here,} all variables are jointly queried. 
   
   {\color{black}Although there was some} some volume effect allowing for reasonable results, {\color{black} judicious choice of the query candidate had a significant impact on the overall quality of the result. To demonstrate this observation, in} Section \ref{subsec:randomActive}, {\color{black}we compare the} RMS error and Std.Dev. for active sampling and for random sampling. 
   
   {\color{black}Ultimately, the amount of time that} can be saved in the engineering design is a function of the uncertainty {\color{black} that can be tolerated} in the inference process. If no error can be tolerated, such as in a critical validation phase, all simulations must be run. If some error can be tolerated, as is the case in a intermediary step, it is possible to use the regression process to skip uninformative simulations. Therefore the potential speedup can only be assessed from the analysis of the convergence of the inference process.

   \subsection{Single variable selection}\label{subsec:singleVar}
   
     \hspace{1em} To exemplify and explore the effect of the methodology presented in this work, we ran the standard uncertainty sampling over the values of DNVUF-201 and FX variables for the three filling cases, {\color{black} namely} empty, water and mixture of water and oil (mean). This provides a baseline case for how active learning works in our problem. It also allows us to have a reference {\color{black}for} the convergence curves in the simplest case, {\color{black} against which the results presented in Section \ref{subsec:multiVar} can be compared}.
     
     The values we {\color{black} computed were} the RMS error $\epsilon_{{\rm RMS},\mathcal{D}_n} $, the maximum error $\epsilon_{{\rm Max},\mathcal{D}_n}$, and the corresponding implied standard deviations $\sigma_{{\rm RMS},\mathcal{D}_n}$ and $\sigma_{{\rm Max},\mathcal{D}_n}$ over $\mathcal{U}$.
     
\begin{figure}[H]        
        \begin{minipage}[b]{0.49\linewidth}
          \includegraphics[width=\linewidth]{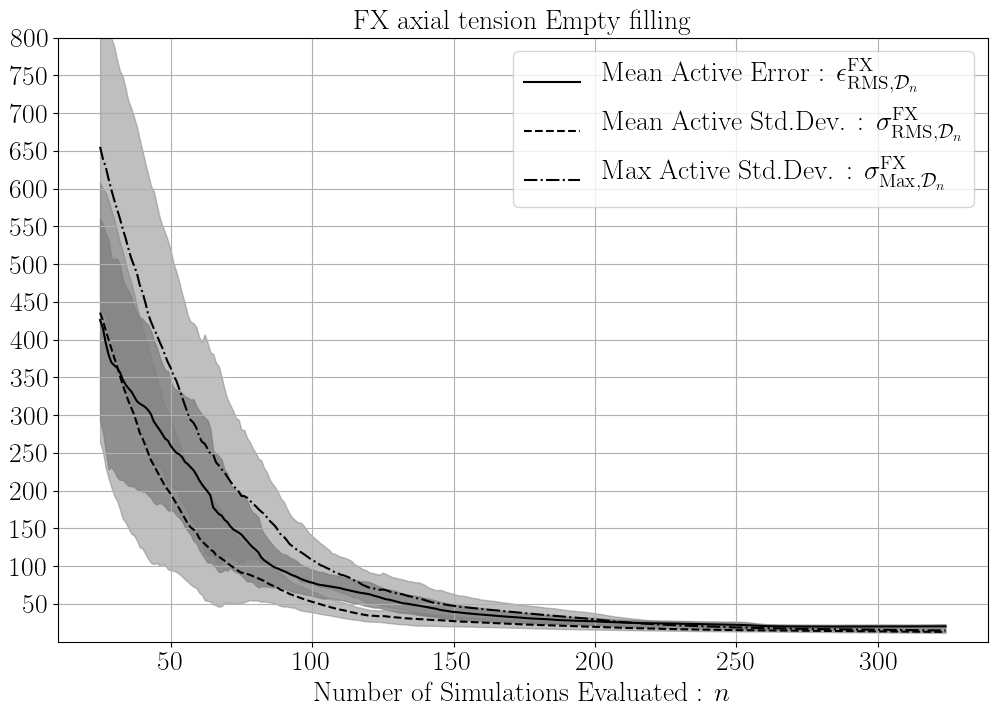} 
        \end{minipage}
        \begin{minipage}[b]{0.49\linewidth}
          \includegraphics[width=\linewidth]{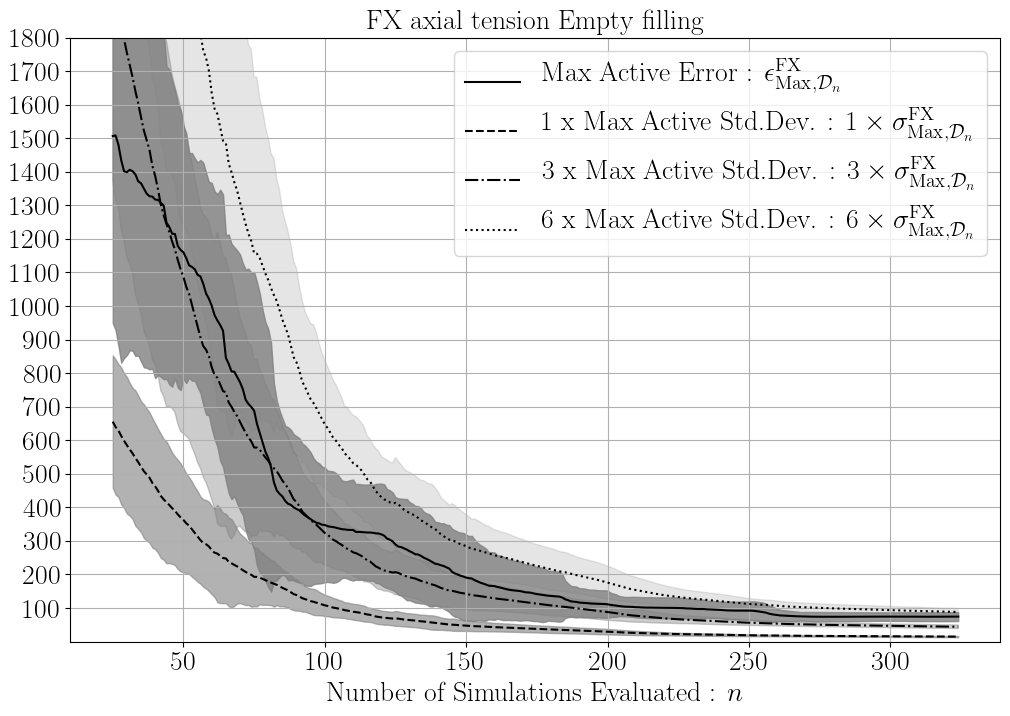}    
        \end{minipage}
        
        \begin{minipage}[b]{0.49\linewidth}
          \includegraphics[width=\linewidth]{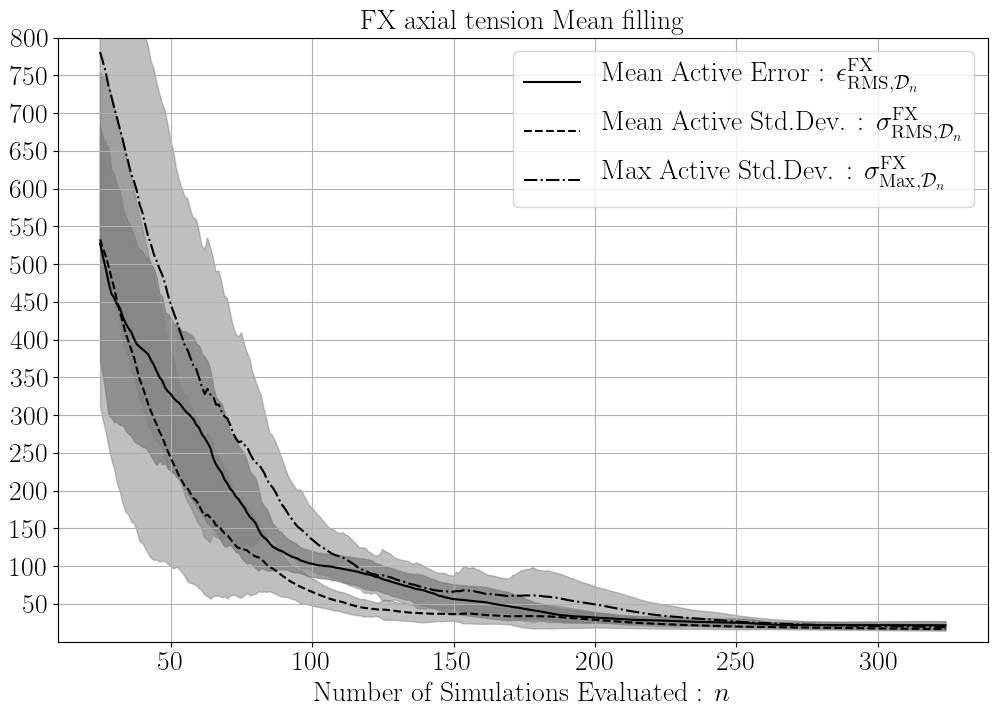} 
        \end{minipage}
        \begin{minipage}[b]{0.49\linewidth}
          \includegraphics[width=\linewidth]{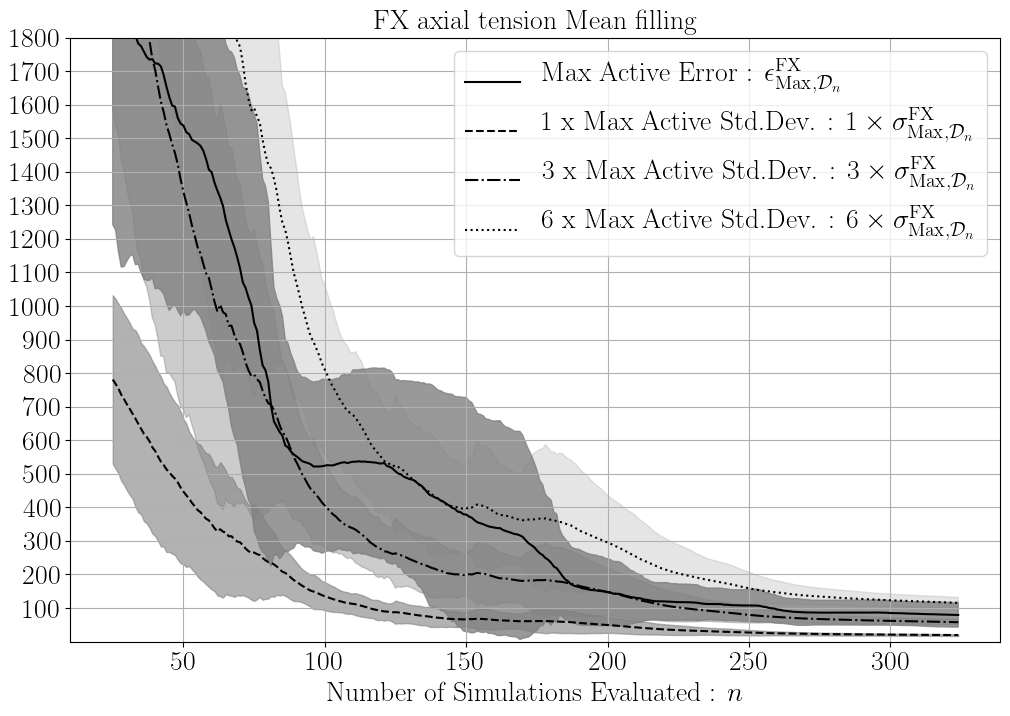}    
        \end{minipage}
        
        \begin{minipage}[b]{0.49\linewidth}
          \includegraphics[width=\linewidth]{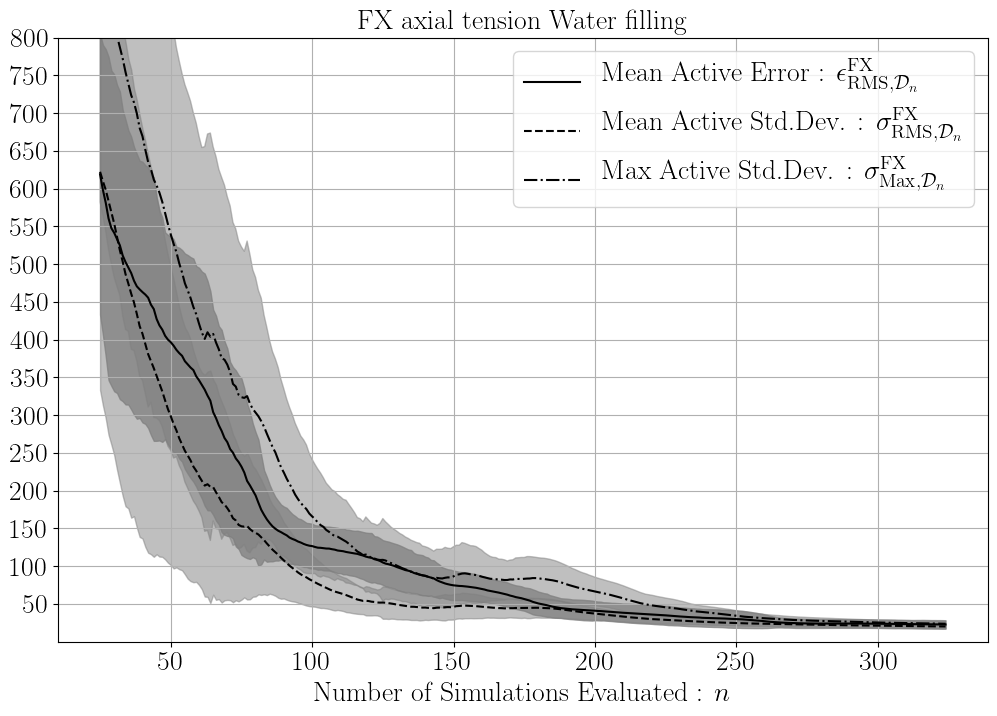} 
        \end{minipage}
        \begin{minipage}[b]{0.49\linewidth}
          \includegraphics[width=\linewidth]{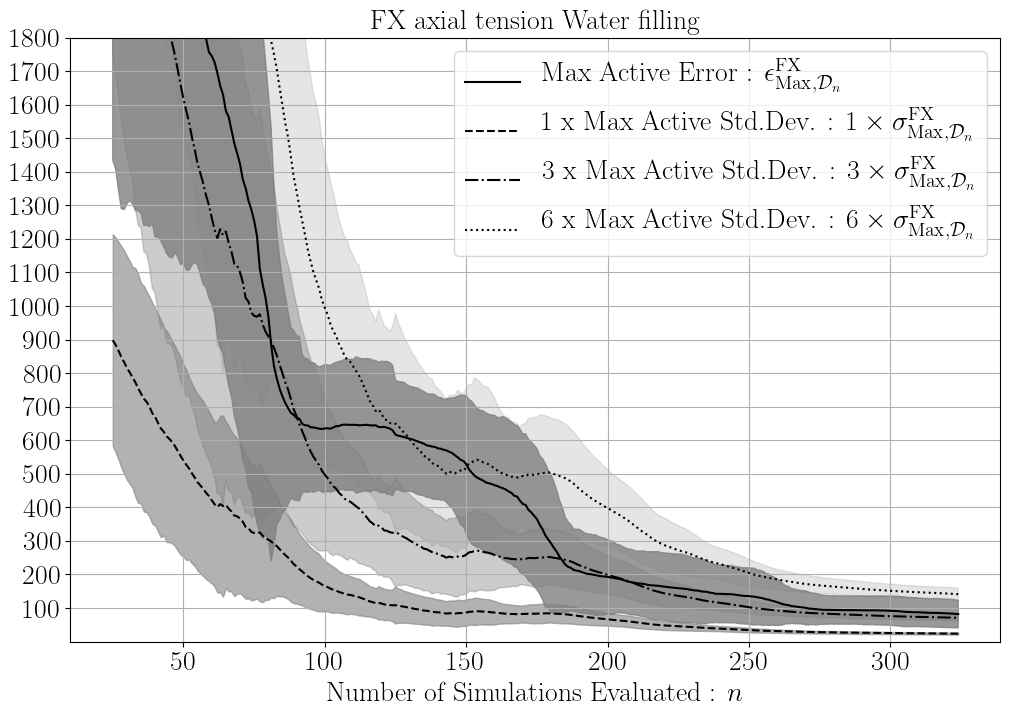}    
        \end{minipage}
        \caption{Measures of deviation for {\color{black}the predictions of} FX in relation $n$. {\color{black}The left-side plots} show $\epsilon_{{\rm RMS},\mathcal{D}_n}^{\rm FX}$, $\sigma_{{\rm RMS},\mathcal{D}_n}^{\rm FX}$ and $\sigma_{{\rm Max},\mathcal{D}_n}^{\rm FX}$. {\color{black} The right-side plots} show $\epsilon_{{\rm Max},\mathcal{D}_n}^{\rm FX}$ against multiples of $\sigma_{{\rm Max},\mathcal{D}_n}^{\rm FX}$. First line refers to empty, the second to mean and third to Water fillings. {\color{black} The curves} correspond to average over all {\color{black} the} seeds and configurations, and shaded areas correspond to 95\% confidence intervals. Sampling was done separately for each variable.}\label{fig:singleTargetFX}
      \end{figure}

     {\color{black} Furthermore, all the} results were based on $6$ different riser configurations and $n=33$ runs with different seeds for the initial sampling, which we averaged {\color{black} over} to produce the curves. {\color{black}In addition, we computed the} 95\% confidence intervals, with each variable being sampled alone. The results can be seen in Figures \ref{fig:singleTargetDNV} and \ref{fig:singleTargetFX}:
     
     \begin{figure}[H]        
        \begin{minipage}[b]{0.49\linewidth}
          \includegraphics[width=\linewidth]{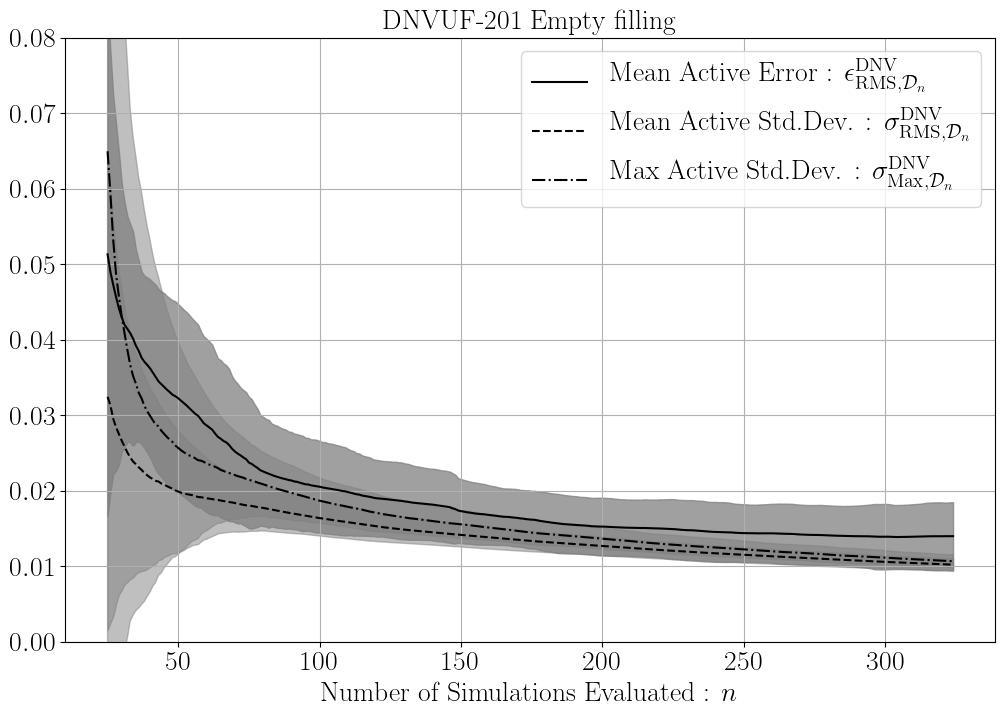} 
        \end{minipage}
        \begin{minipage}[b]{0.49\linewidth}
          \includegraphics[width=\linewidth]{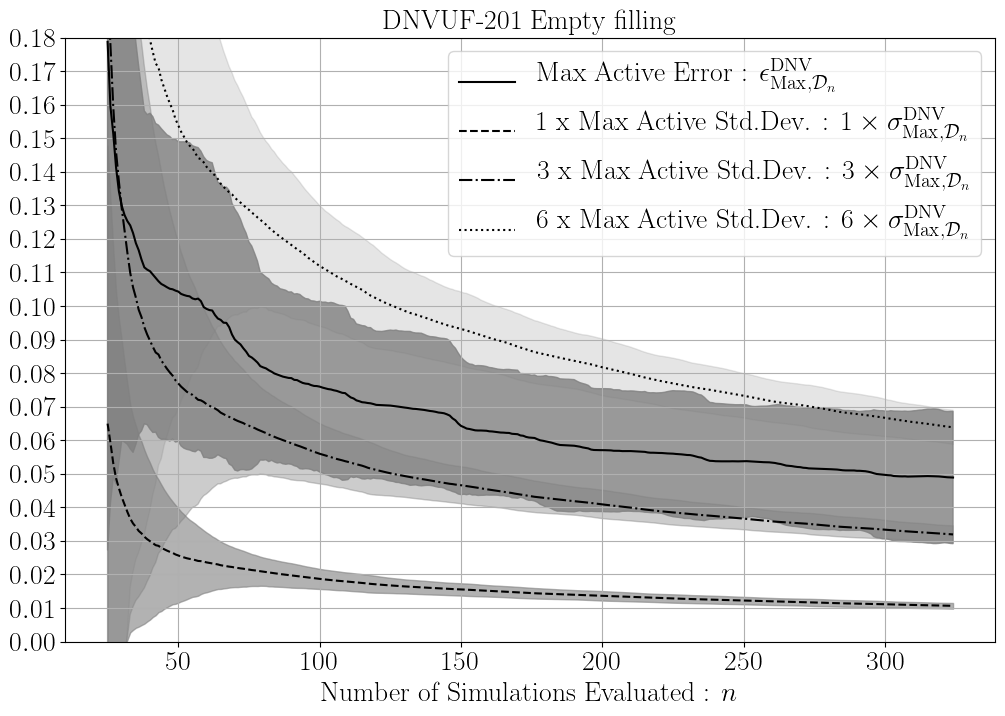}    
        \end{minipage}
        
        \begin{minipage}[b]{0.49\linewidth}
          \includegraphics[width=\linewidth]{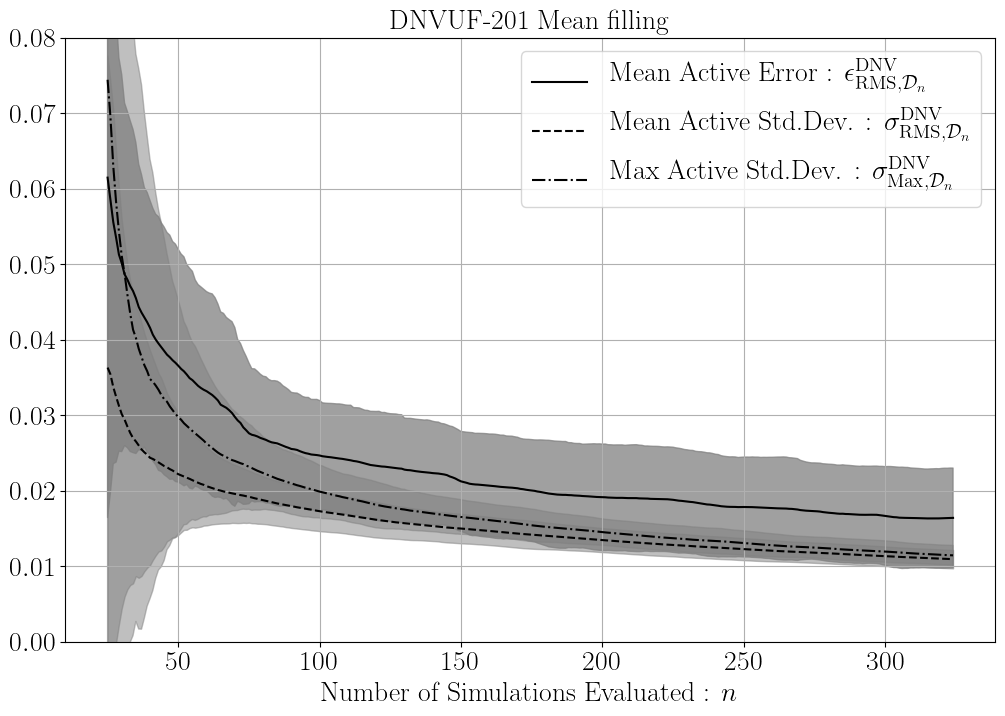} 
        \end{minipage}
        \begin{minipage}[b]{0.49\linewidth}
          \includegraphics[width=\linewidth]{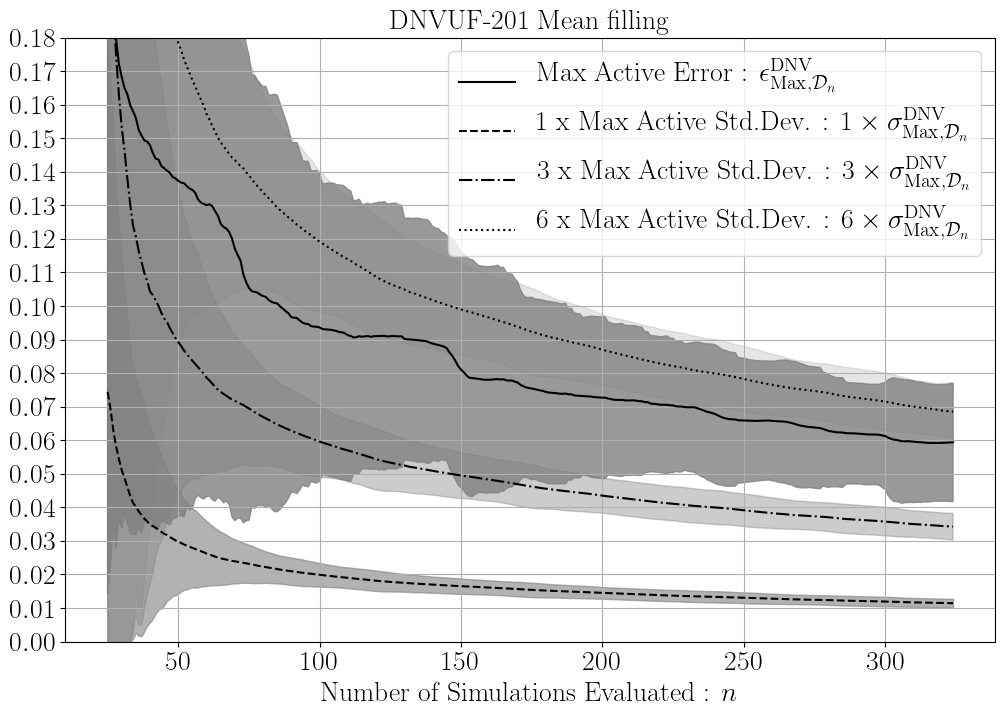}    
        \end{minipage}
        
        \begin{minipage}[b]{0.49\linewidth}
          \includegraphics[width=\linewidth]{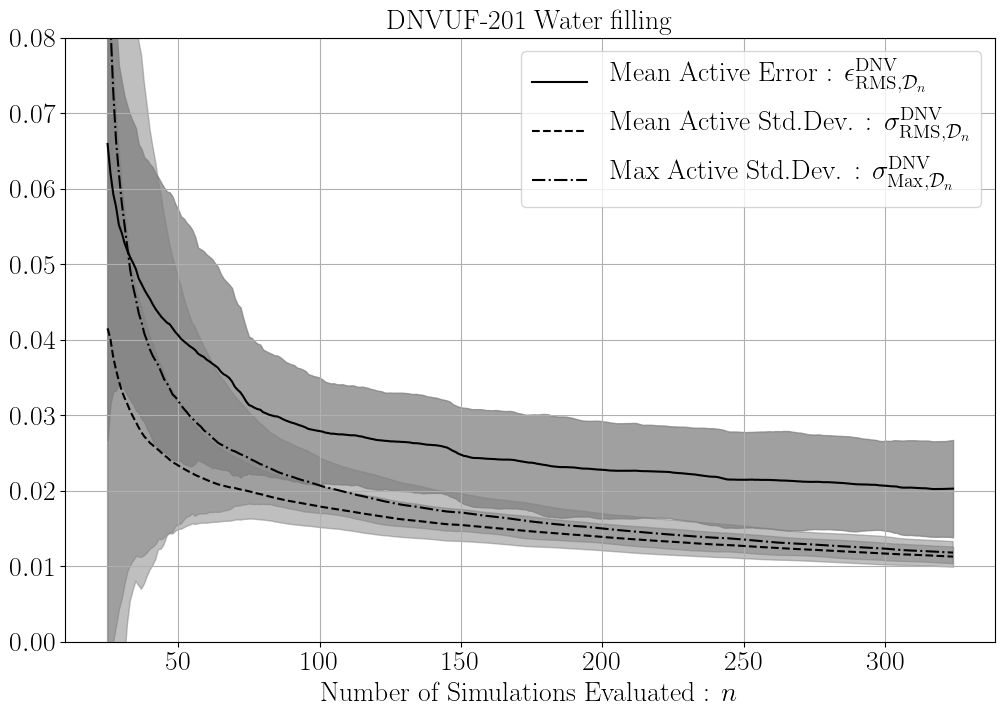} 
        \end{minipage}
        \begin{minipage}[b]{0.49\linewidth}
          \includegraphics[width=\linewidth]{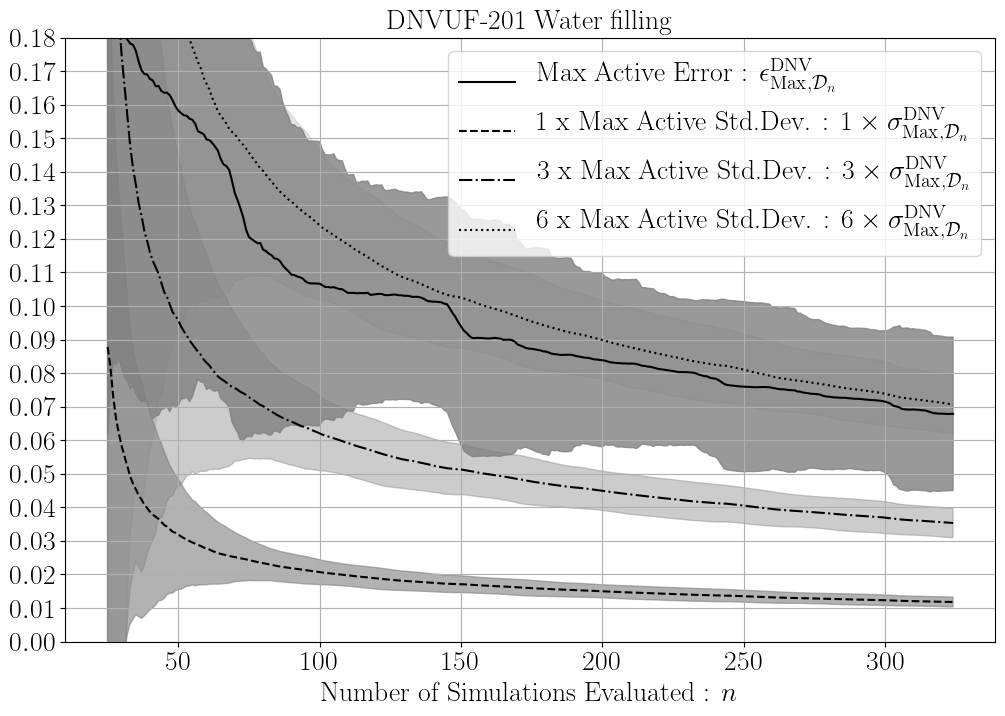}    
        \end{minipage}
        \caption{Measures of deviation for {\color{black}the predictions of} DNVUF-201 in relation $n$. {\color{black}The left-side plots} show $\epsilon_{{\rm RMS},\mathcal{D}_n}^{\rm DNV}$, $\sigma_{{\rm RMS},\mathcal{D}_n}^{\rm DNV}$ and $\sigma_{{\rm Max},\mathcal{D}_n}^{\rm DNV}$. {\color{black} The right-side plots} show $\epsilon_{{\rm Max},\mathcal{D}_n}^{\rm DNV}$ against multiples of $\sigma_{{\rm Max},\mathcal{D}_n}^{\rm DNV}$. First line refers to empty, the second to mean and third to Water fillings. {\color{black} The curves} correspond to average over all {\color{black} the} seeds and configurations, and shaded areas correspond to 95\% confidence intervals. Sampling was done separately for each variable.}\label{fig:singleTargetDNV}
      \end{figure}
     
     {\color{black}From the left-side graphs} in Figures \ref{fig:singleTargetDNV} and \ref{fig:singleTargetFX}, we can observe that the RMS {\color{black} error decreased monotonically, as the number of samples increased}. {\color{black}Furthermore, we} observe that the RMS error {\color{black}could} be reasonably bounded by $\epsilon_{{\rm RMS},\mathcal{D}_n}^{\rm DNV} < \sigma_{{\rm RMS},\mathcal{D}_n}^{\rm DNV} + \sigma_0$ with $0.1 \leq \sigma_0 \leq 0.15$, and $\epsilon_{{\rm Max},\mathcal{D}_n}^{\rm DNV} < k \sigma_{{\rm Max},\mathcal{D}_n}^{\rm DNV}$ for $4.5 \leq k \leq 6$. For axial tension, $\epsilon_{{\rm RMS},\mathcal{D}_n}^{\rm FX} \approx (\sigma_{{\rm RMS},\mathcal{D}_n}^{\rm FX} + \sigma_{{\rm Max},\mathcal{D}_n}^{\rm FX})/2$, while being mostly bounded by $\sigma_{{\rm Max},\mathcal{D}_n}^{\rm FX}$. Also $\epsilon_{{\rm Max},\mathcal{D}_n}^{\rm FX} < k \sigma_{{\rm Max},\mathcal{D}_n}^{\rm FX}$ for $6.0 \leq k \leq 9.0$. 
          
     It is also interesting to notice that the confidence intervals for the implied Std.Dev. become reasonably tight after 50 to 100 evaluations, {\color{black} implying} that the results produced by the active learning loop are fairly consistent and robust for different choices of initial sample. This supports the hypothesis of active learning as producing optimal query sequence to maximize information contained in the {\color{black}labeled} dataset $\mathcal{D}_n$. 
        
     Considering reasonable RMS errors $\epsilon_{{\rm RMS},\mathcal{D}_n}^{\rm DNV} < 0.025$ and for Axial Tension around $\epsilon_{{\rm RMS},\mathcal{D}_n}^{FX} < 200\ {\rm N}$, it is possible to obtain an inference over the entire unlabeled dataset $\mathcal{U}$ with only 100 evaulations, while also keeping the maximum error $\epsilon_{{\rm Max},\mathcal{D}_n}^{DNV} < 0.10$ and $\epsilon_{{\rm Max},\mathcal{D}_n}^{FX} < 1000\ {\rm N}$, which is a reasonable tolerance in the context of these simulations. 
      
   \subsection{Multi-variable aggregate selection}\label{subsec:multiVar}
   
     \hspace{1em} {\color{black} The primary interest, as explained in Section \ref{sec:Intro}, was to be able to run smart sequential simulations that return multiple results, so that values of all the variables produced by the simulation can be jointly inferred in the most efficient way possible. To do so, we adopted the standard uncertainty sampling, using the geometric mean of the uncertainties of all six target values as uncertainty. These were the values of DNVUF-201 and FX for the 3 filling cases, empty, water and mixture of water and oil (mean).}
     
\begin{figure}[H]        
        \begin{minipage}[b]{0.49\linewidth}
          \includegraphics[width=\linewidth]{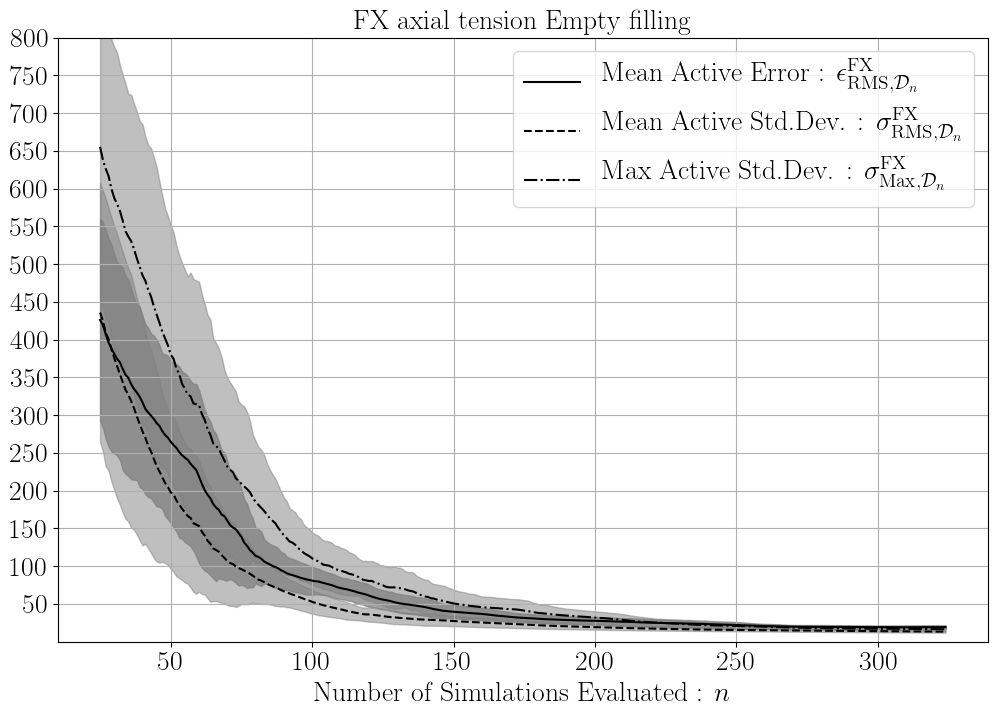} 
        \end{minipage}
        \begin{minipage}[b]{0.49\linewidth}
          \includegraphics[width=\linewidth]{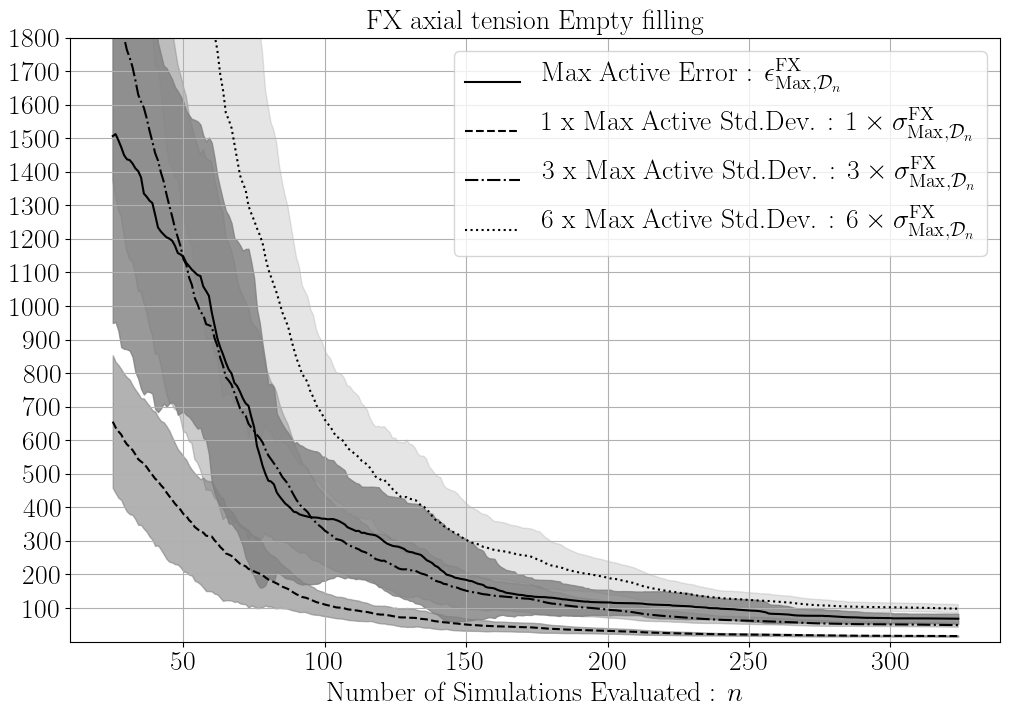}    
        \end{minipage}
        
        \begin{minipage}[b]{0.49\linewidth}
          \includegraphics[width=\linewidth]{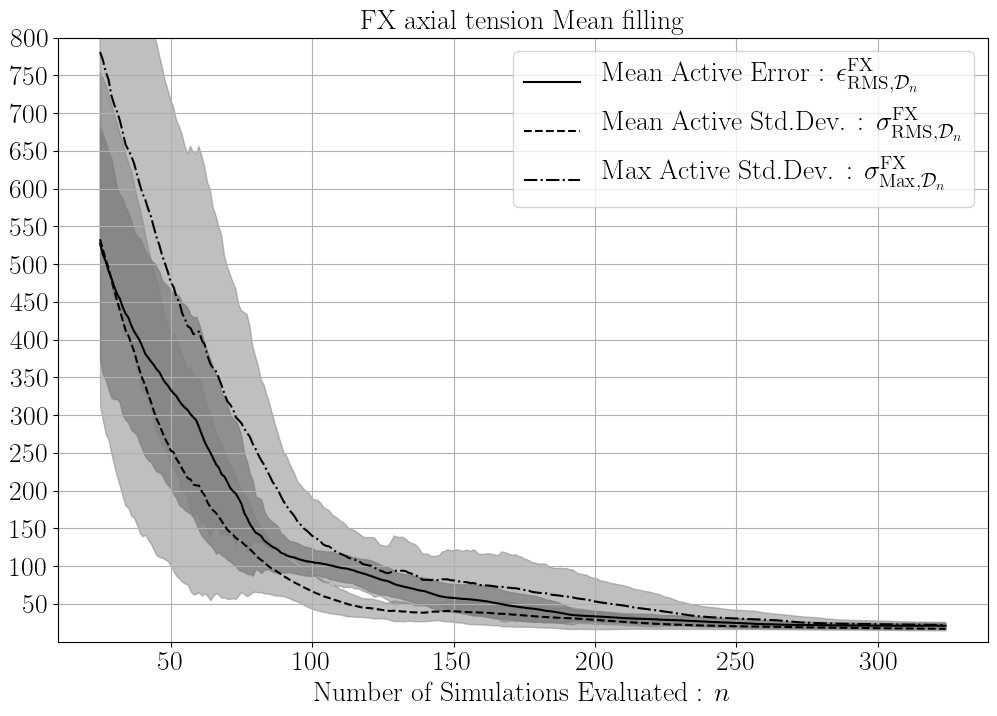} 
        \end{minipage}
        \begin{minipage}[b]{0.49\linewidth}
          \includegraphics[width=\linewidth]{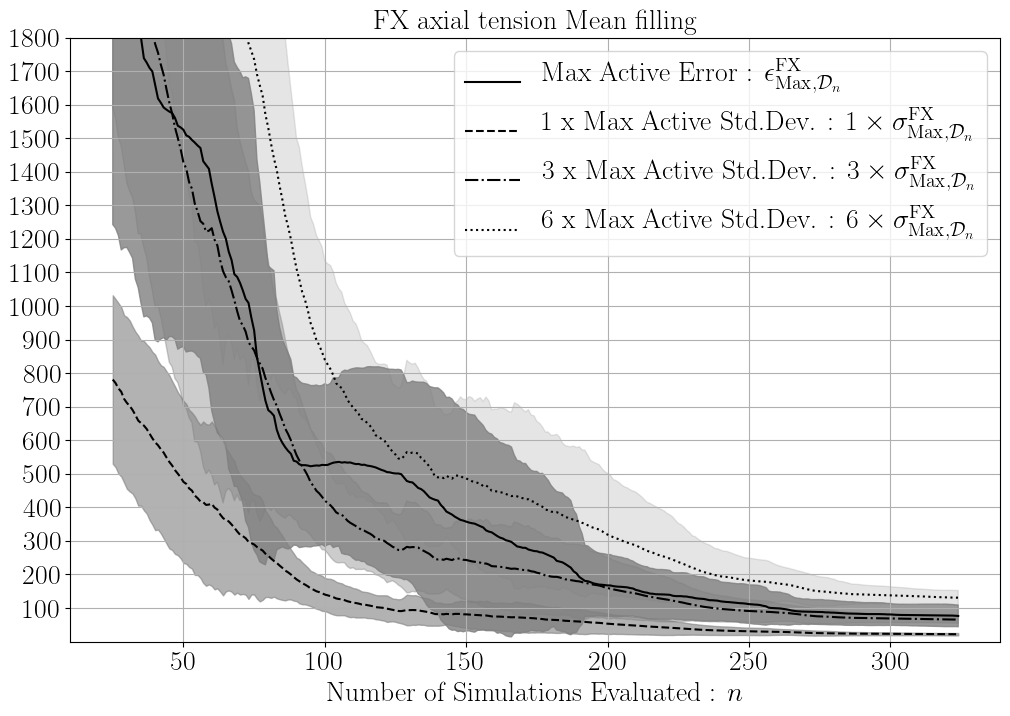}    
        \end{minipage}
        
        \begin{minipage}[b]{0.49\linewidth}
          \includegraphics[width=\linewidth]{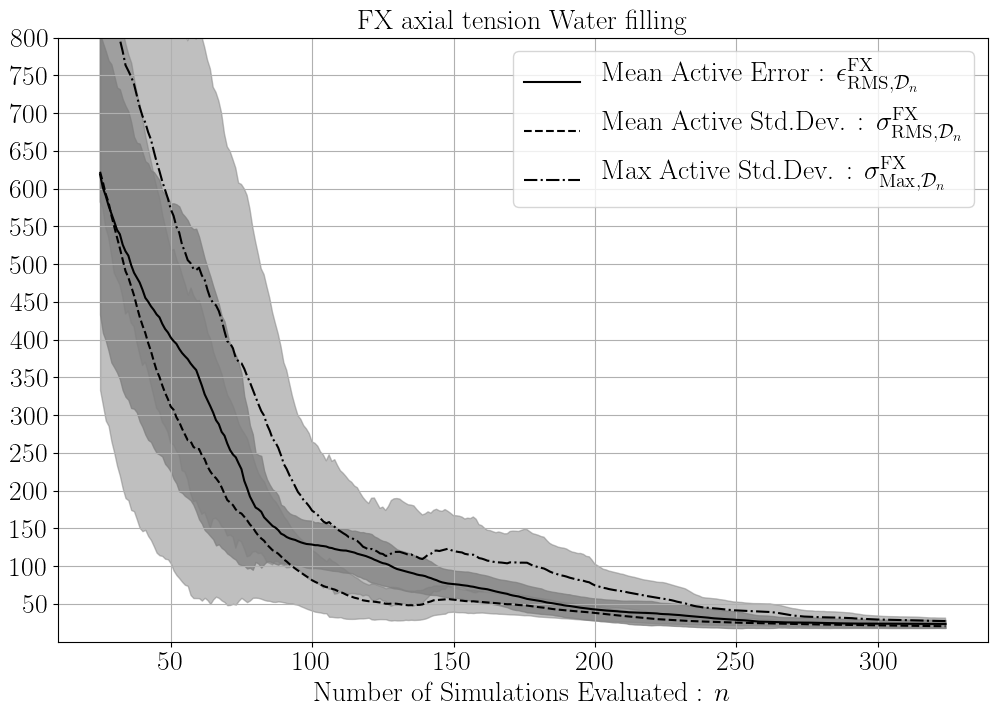} 
        \end{minipage}
        \begin{minipage}[b]{0.49\linewidth}
          \includegraphics[width=\linewidth]{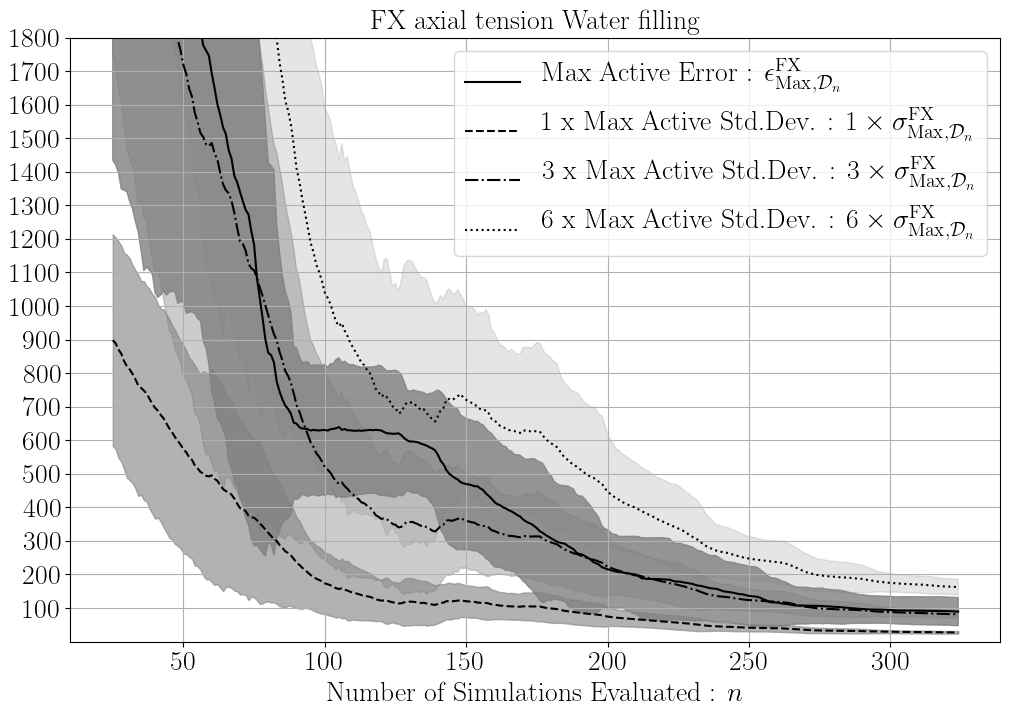}    
        \end{minipage}
        \caption{Measures of deviation for {\color{black}the predictions of} FX in relation $n$. {\color{black}The left-side plots} show $\epsilon_{{\rm RMS},\mathcal{D}_n}^{\rm FX}$, $\sigma_{{\rm RMS},\mathcal{D}_n}^{\rm FX}$ and $\sigma_{{\rm Max},\mathcal{D}_n}^{\rm FX}$. {\color{black} The right-side plots} show $\epsilon_{{\rm Max},\mathcal{D}_n}^{\rm FX}$ against multiples of $\sigma_{{\rm Max},\mathcal{D}_n}^{\rm FX}$. First line refers to empty, the second to mean and third to Water fillings. {\color{black} The curves} correspond to average over all {\color{black} the} seeds and configurations, and shaded areas correspond to 95\% confidence intervals. Sampling was done separately for each variable. }\label{fig:multiTargetFX}
      \end{figure}

     In this case, the error values are done for all variables on a single sampling sequence, with the points used to compute the inference over the same {\color{black} set} $\mathcal{U}$ for all target variables. The results are presented in the same format as in Section \ref{subsec:singleVar}. {\color{black} However}, they represent represent the results of the joint sampling of all variables. All results were based on $n_c=6$ different riser configurations and $n=33$ runs with different seeds for  the initial sampling, which {\color{black} were then averaged over } to produce the curves and also compute 95\% confidence intervals. The results for this case can be seen in Figures \ref{fig:multiTargetDNV} and \ref{fig:multiTargetFX}:

     \begin{figure}[H]        
        \begin{minipage}[b]{0.49\linewidth}
          \includegraphics[width=\linewidth]{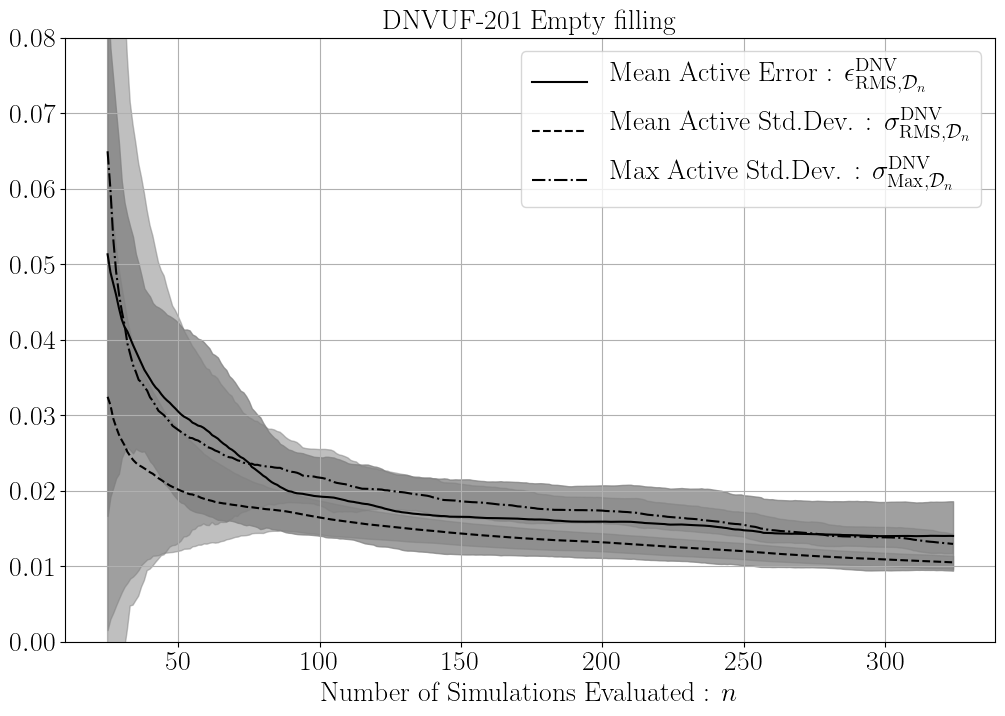} 
        \end{minipage}
        \begin{minipage}[b]{0.49\linewidth}
          \includegraphics[width=\linewidth]{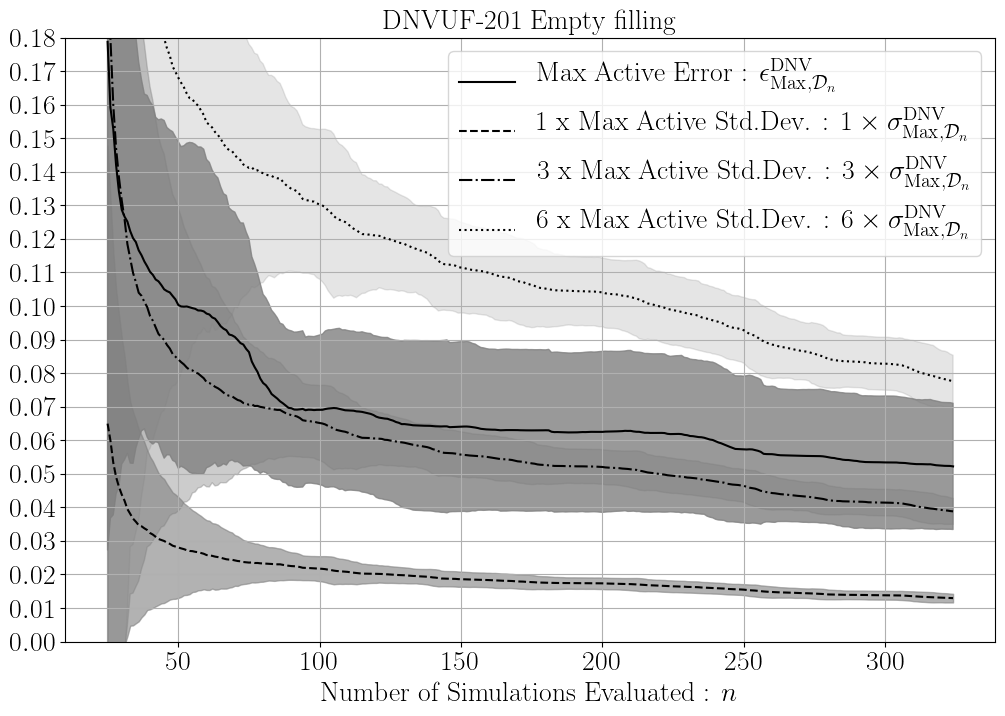}    
        \end{minipage}
        
        \begin{minipage}[b]{0.49\linewidth}
          \includegraphics[width=\linewidth]{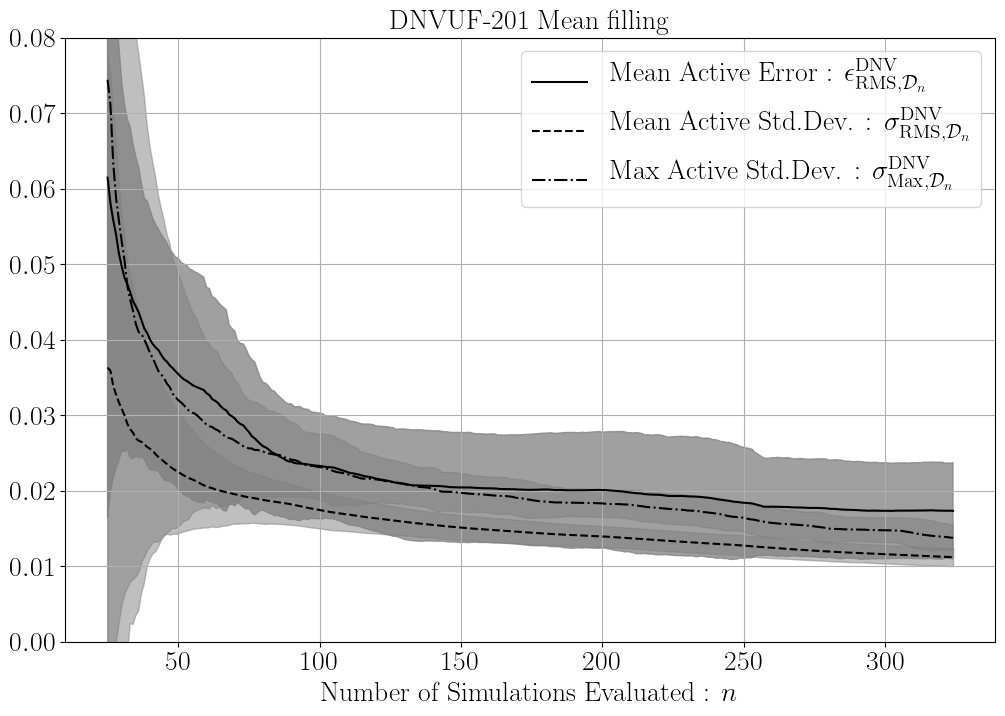} 
        \end{minipage}
        \begin{minipage}[b]{0.49\linewidth}
          \includegraphics[width=\linewidth]{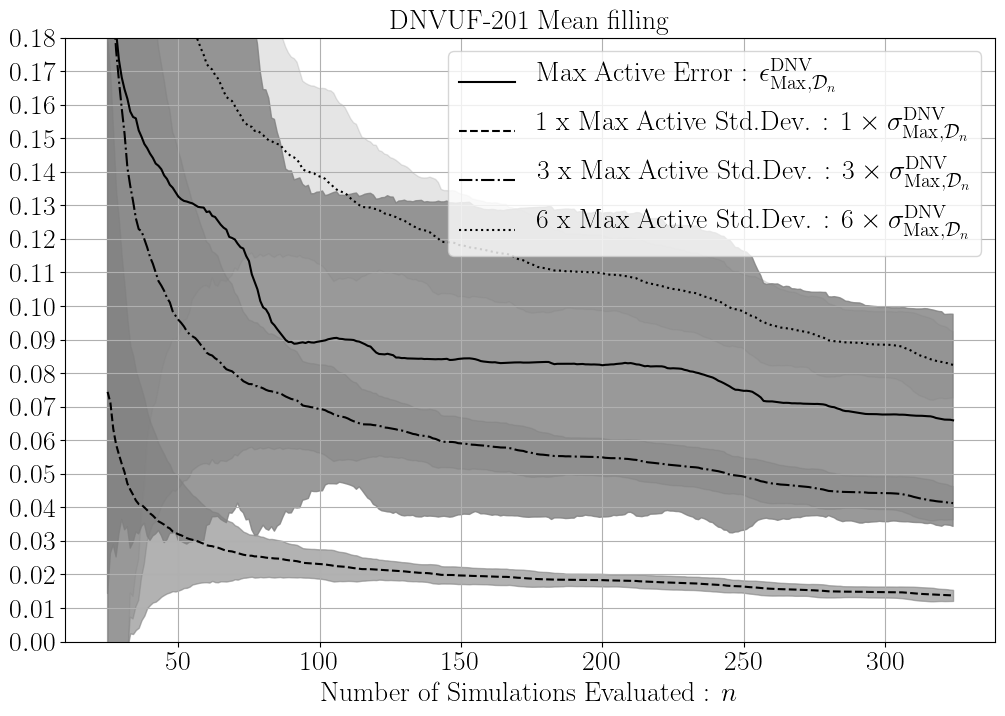}    
        \end{minipage}
        
        \begin{minipage}[b]{0.49\linewidth}
          \includegraphics[width=\linewidth]{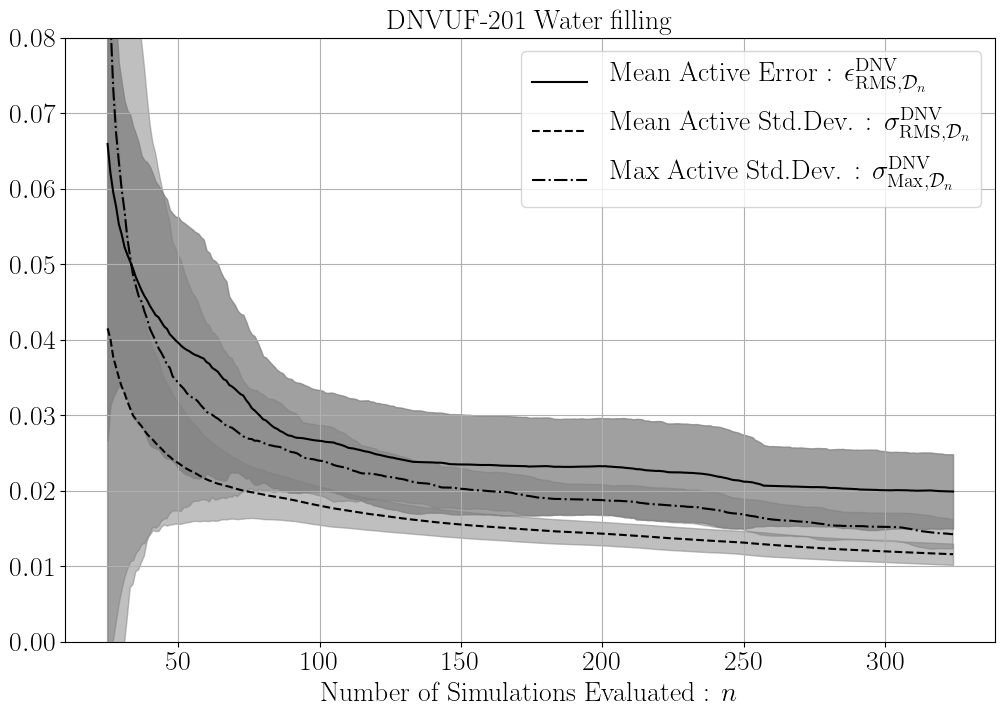} 
        \end{minipage}
        \begin{minipage}[b]{0.49\linewidth}
          \includegraphics[width=\linewidth]{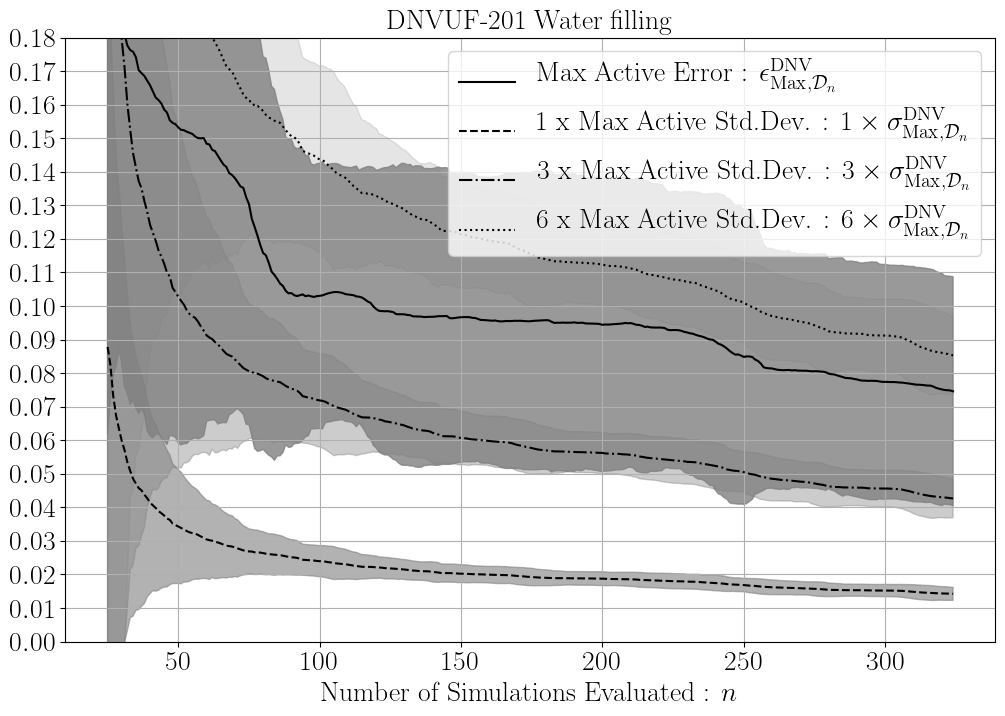}    
        \end{minipage}
        \caption{Measures of deviation for {\color{black}the predictions of} DNVUF-201 in relation $n$. {\color{black}The left-side plots} show $\epsilon_{{\rm RMS},\mathcal{D}_n}^{\rm DNV}$, $\sigma_{{\rm RMS},\mathcal{D}_n}^{\rm DNV}$ and $\sigma_{{\rm Max},\mathcal{D}_n}^{\rm DNV}$. {\color{black} The right-side plots} show $\epsilon_{{\rm Max},\mathcal{D}_n}^{\rm DNV}$ against multiples of $\sigma_{{\rm Max},\mathcal{D}_n}^{\rm DNV}$. First line refers to empty, the second to mean and third to Water fillings. {\color{black} The curves} correspond to average over all {\color{black} the} seeds and configurations, and shaded areas correspond to 95\% confidence intervals. Sampling was done separately for each variable. }\label{fig:multiTargetDNV}
      \end{figure}
     
     {\color{black} A} majority of the results obtained in Section \ref{subsec:singleVar} remain valid, as can be seen in Figures  \ref{fig:multiTargetDNV} and \ref{fig:multiTargetFX}. We continue to have the monotonic decay of the RMS error, and the mostly monotonic decay of the maximum errors. {\color{black} Furthermore,} we continue to have the ability to approximate and bound those errors. {\color{black} In addition, a} majority of the bounds still apply as is in this case, therefore showing that the performance haven't been too degraded by changing from querying each variable separately to querying them together. 
     
     As mentioned in Section \ref{subsec:singleVar}, considering reasonable RMS errors for DNVUF-201 with only 100 evaulations, while also keeping the maximum error under control over all yet-to-be run simulations {\color{black} imply that these results represent} a reduction of {\color{black} more than} 80\% of the number of simulations required, compared to the total  {\color{black} number} of simulations, which was 512. If {\color{black} more relaxed} tolerances {\color{black} could} be accepted, such as  $\epsilon_{{\rm RMS},\mathcal{D}_n}^{\rm DNV} < 0.04$ and  $\epsilon_{{\rm RMS},\mathcal{D}_n}^{\rm DNV} < 400\ {\rm N}$ , it {\color{black} would be possible to} achieve reductions of up to 90\% {\color{black} in} the total number of simulations, while keeping $\epsilon_{{\rm Max},\mathcal{D}_n}^{\rm DNV} < 0.15$ and $\epsilon_{{\rm Max},\mathcal{D}_n}^{\rm FX} < 1500$. {\color{black} Although these criteria appear to be high, they are not completely unsuitable for a quick analysis, } corresponding to approximately 30\% in the worst case scenario.

   \subsection{Random and Active sampling comparison}\label{subsec:randomActive}
   
     \hspace{1em} To evaluate the impact of optimal querying, we compare the results for jointly querying all variables together for both uncertainty sampling and for random sampling. {\color{black} Each } random sampling case corresponds an uncertainty sampling case with the same initial points. At each step, the we randomly sample without replacement a case from $\mathcal{U}$, otherwise it is the same as described in Section \ref{subsec:multiVar}. Similar results can be obtained for the sampling performed in Section \ref{subsec:singleVar}. The results can be seen in Figures \ref{fig:randomActive}:
     
     \begin{figure}[H]        
        \begin{minipage}[b]{0.49\linewidth}
          \includegraphics[width=\linewidth]{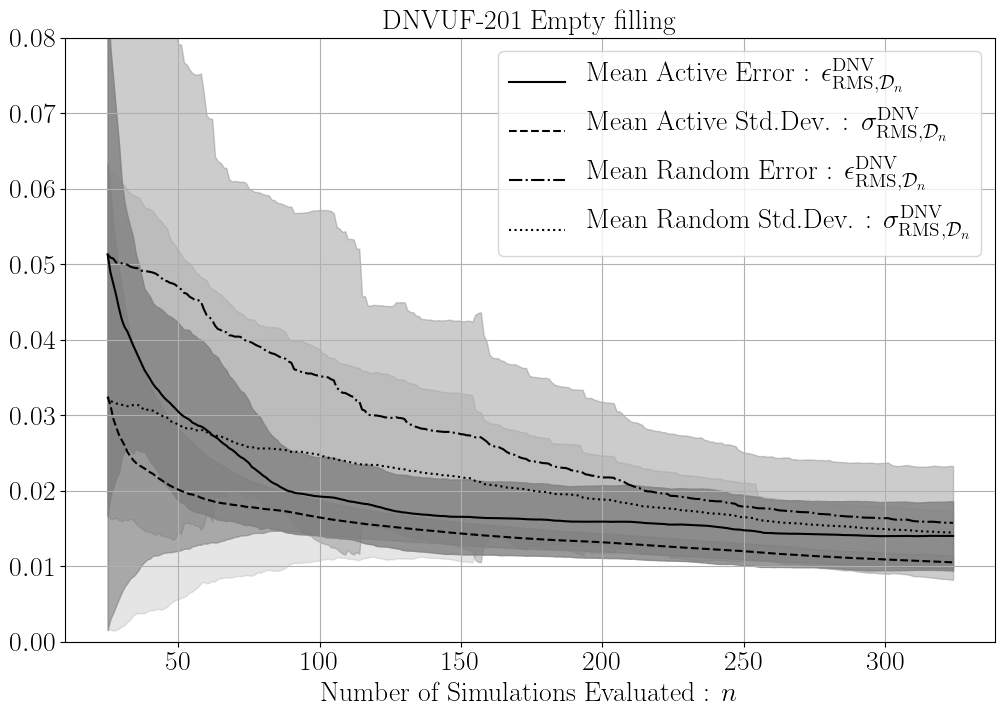} 
        \end{minipage}
        \begin{minipage}[b]{0.49\linewidth}
          \includegraphics[width=\linewidth]{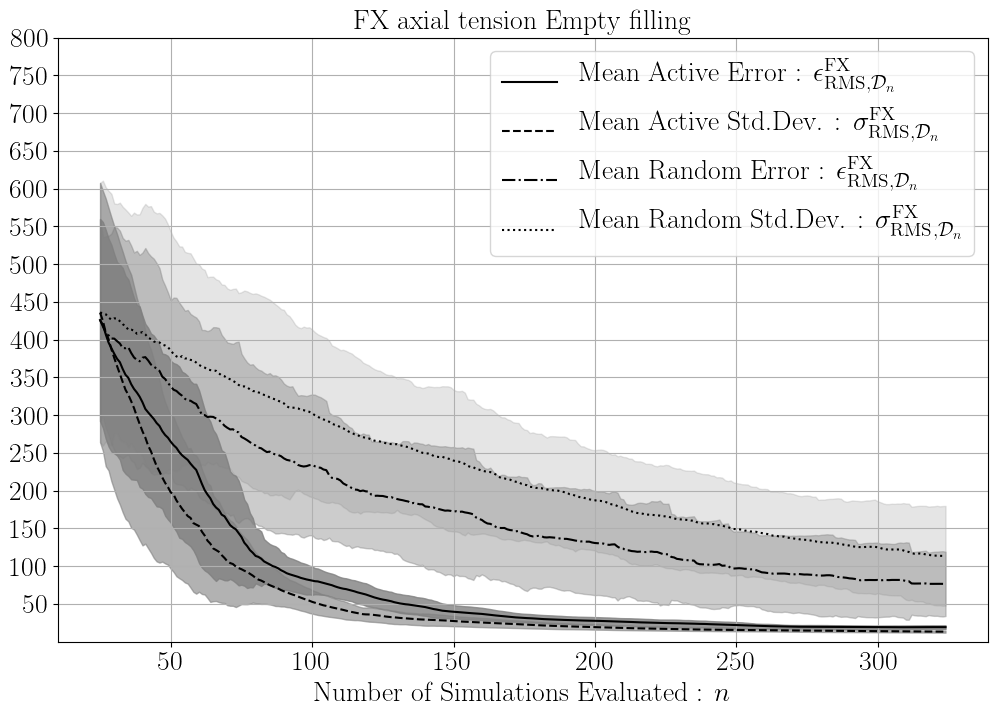} 
        \end{minipage}
        
        \begin{minipage}[b]{0.49\linewidth}
          \includegraphics[width=\linewidth]{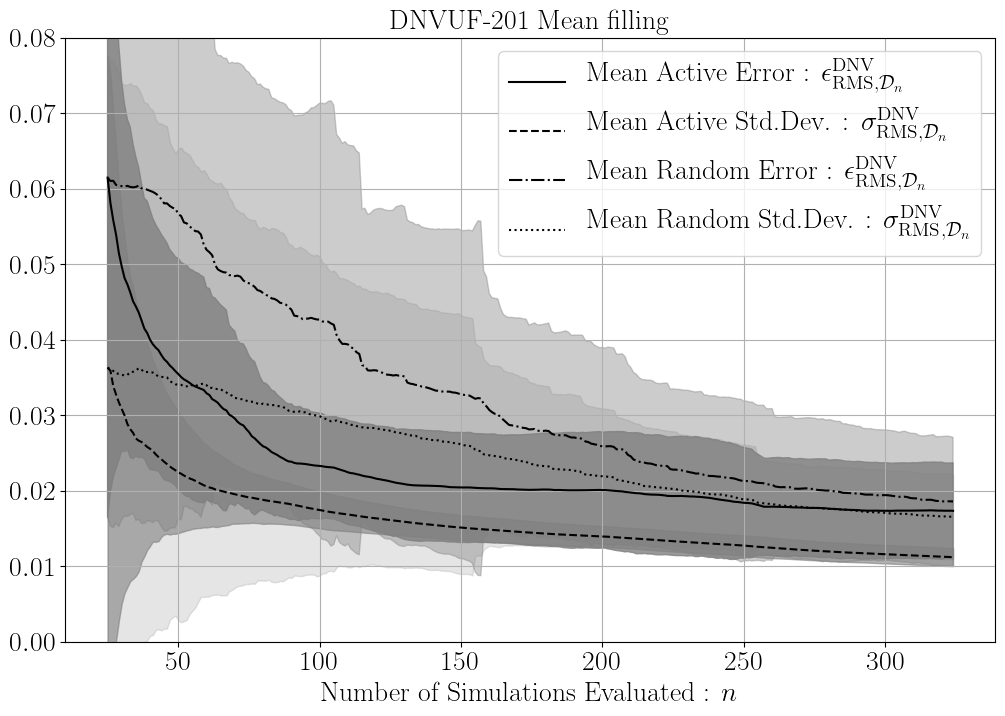} 
        \end{minipage}
        \begin{minipage}[b]{0.49\linewidth}
          \includegraphics[width=\linewidth]{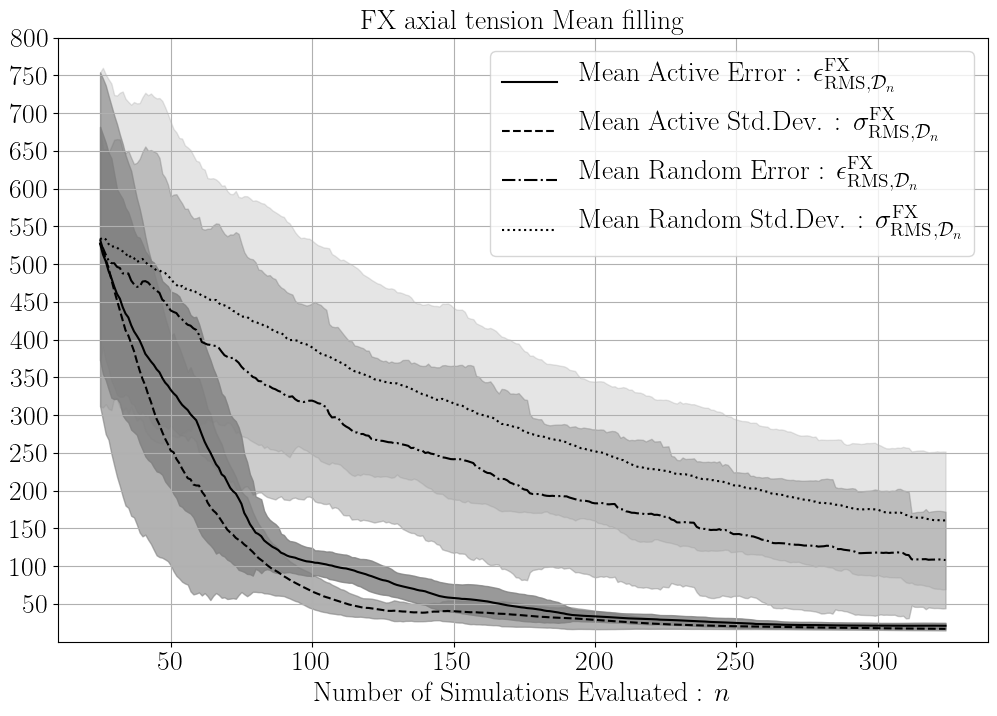} 
        \end{minipage}
        
        \begin{minipage}[b]{0.49\linewidth}
          \includegraphics[width=\linewidth]{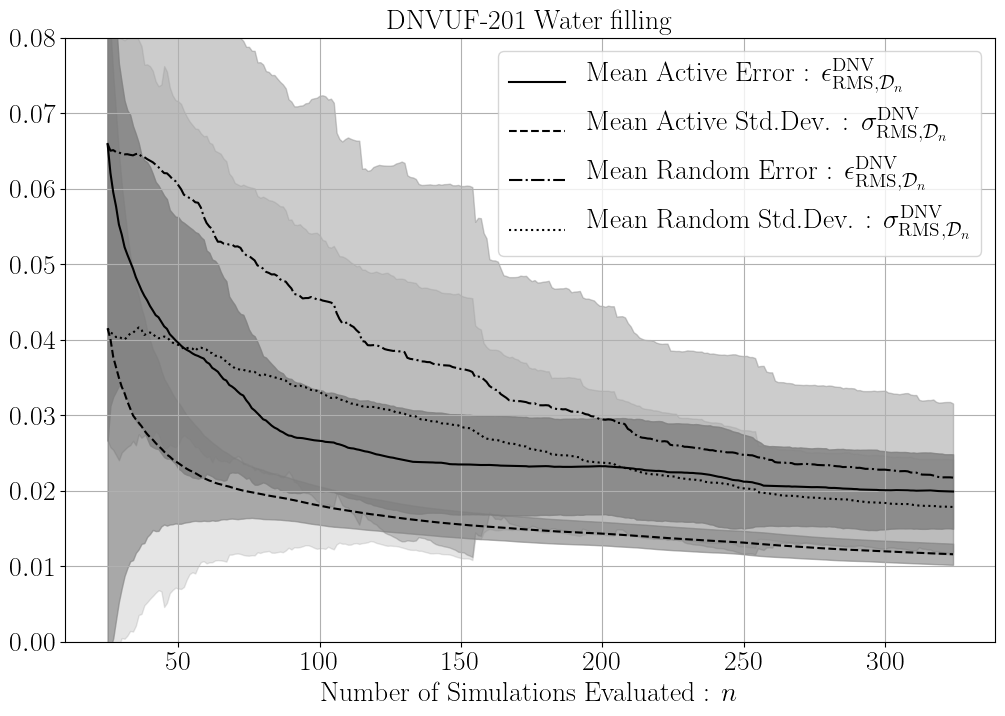} 
        \end{minipage}
        \begin{minipage}[b]{0.49\linewidth}
          \includegraphics[width=\linewidth]{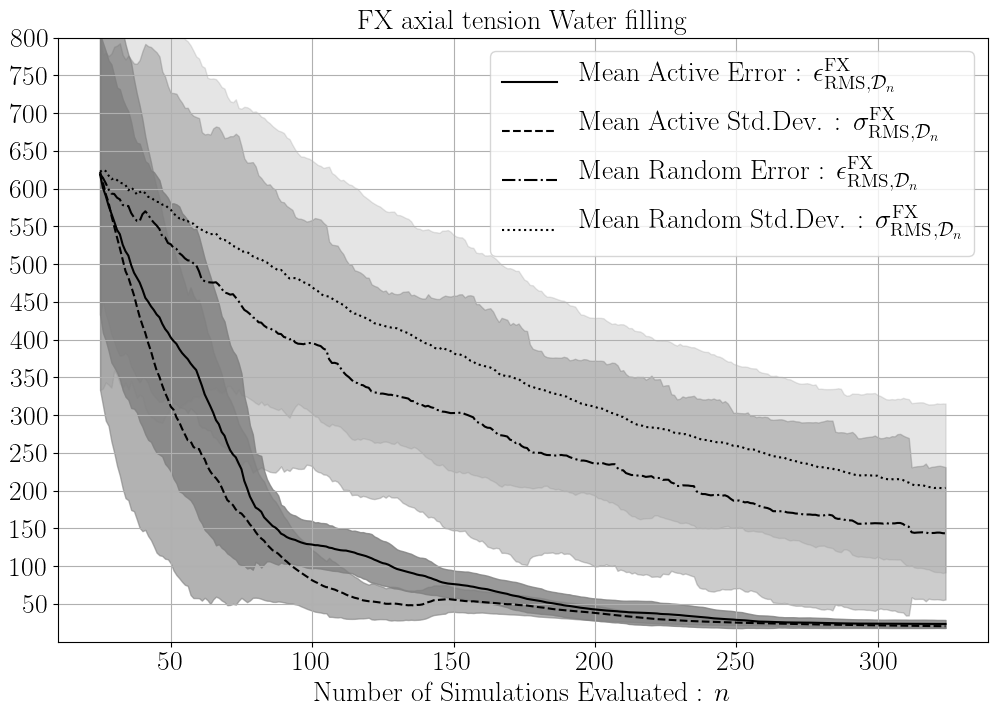} 
        \end{minipage}
        \caption{Measures of deviation for DNVUF-201 and ${\rm FX}$ predictions in relation to $n$. {\color{black}The left-side plots show $\epsilon_{{\rm RMS},\mathcal{D}_n}^{\rm DNV}$ and $\sigma_{{\rm RMS},\mathcal{D}_n}^{\rm DNV}$ for the active learning and random samplings for DNVUF-201. The right side plots show $\epsilon_{{\rm RMS},\mathcal{D}_n}^{\rm FX}$ and $\sigma_{{\rm RMS},\mathcal{D}_n}^{\rm FX}$ for the active learning and random samplings for FX. The first line refers to empty case, while the second and third lines refer to the mixture and water filled cases. The curves correspond to the average over all the seeds and configurations, and shaded areas correspond to 95\% confidence intervals. Sampling was performed jointly for all variables.} }\label{fig:randomActive}
      \end{figure}
      
      The most notable feature of the results {\color{black} was} that the average error for random sampling have not only larger $\epsilon_{{\rm RMS},\mathcal{D}_n} $, but much larger spread than the active learning results. This is not unexpected since it reasonable to consider that an efficient sampling criteria will be more {\color{black}consistent} than {\color{black} random selection}. {\color{black} The much higher spread shows that there was a highly uneven variety of results that could be obtained from the randomly sampled points; however, it was not the case with the uncertainty sampling. This further corroborates the robustness of the results obtained from uncertainty sampling.}
      
      {\color{black} In summary, } it is clear that the average error also decreases with the increasing number, and the error per se {\color{black} was acceptable}, thus allowing for a reasonable approximation of $\mathcal{U}$; {\color{black} however, this result was inferior to that from the uncertainty sampling.} This supports the use of {\color{black} direct random sampling or obtaining some points through a design of experiment, and then feeding them directly to a machine learning model,} as mentioned in Section \ref{sec:Intro}. {\color{black}The } decrease in the RMS error comes purely from an increasing number of evaluations, which is the volume effect mentioned earlier. To obtain results similar to the ones aimed at Section \ref{subsec:multiVar}, one would need upwards of 250 evaluations, with is much less efficient than uncertainty sampling. 
      
%\subsection{Thompson Uncertainty Sampling}

\section{Conclusions} \label{sec:Conclusion}

  \hspace{1em} {\color{black} We validated the} application of the technique of active learning in aiding engineering design {\color{black} of an oil and gas riser, over a set of loading conditions. The proposed method lowered the execution cost required to acquire a reasonable picture of the response of the physical system. We showed that it is feasible to obtain satisfactory results with 80\% fewer simulations than that required by the complete set of cases, thus representing a five-fold reduction in the execution time.}  
  
  We validated the methodology by comparing the results of the inference from the machine learning model against {\color{black} a} reference set of performed simulations. We {\color{black} showed} that the RMS and maximum {\color{black} errors} can be robustly approximated and bounded by affine combinations of RMS and maximum implied uncertainty. This {\color{black} enables} the use of the inference to create usable confidence intervals for the yet-to-be-run simulations, which in turn can be chosen to run or not be run, thus allowing for the mentioned improvement in design time to take place. 
  
  The methodology presented here is agnostic to the origin of the data fed to the machine learning model; therefore, {\color{black} it is} readily applicable to other cases, in which a range of potentially similar simulation must be run to assess the behavior of a physical system in the context of a engineering design problem. The major difference for new cases {\color{black} lies in preparing} the input variables that represent the simulations parameters in a way that is appropriate for the machine learning model to train and infer. 
  
  We also note that classes of engineering design problems, {\color{black} which} require experiments to be run instead of simulations, {\color{black} can} also be addressed by this methodology, because the machine learning model, as mentioned {\color{black} earlier}, {\color{black} is} agnostic to the origin of the data, {\color{black} and the objective is to omit certain expensive, complex, or difficult steps of the design.} This step is performed \textit{in silico} in our case, but could also be done in a laboratory or in a field system.
  
  The performance {\color{black} can}, in principle, be further improved by adjusting the machine learning model, such as by choosing a better covariance kernel in our case. Additionally, parallelization of the sampling could possibly be achieved through some variation of Thompson sampling, {\color{black} enabling} parallel execution of the active learning loop. Finally, alternative forms of normalization or assembling the feature vectors may have additional positive effects on the performance of the algorithm, and this is currently under  investigation by the authors.

\section{Acknowledgements}

  \hspace{1em} This work was partially supported by the National Council for Scientific and Technological Development (CNPq - Conselho Nacional de Desenvolvimento Científico e Tecnológico - Brazil). It was also financed by the Coordenação de Aperfeiçoamento de Pessoal de Nível Superior - Brasil (CAPES) - Finance Code
001.

The authors acknowledge the support provided by the Tecgraf Institute of Technical-Scientific Software Development of PUC-Rio (Tecgraf/PUC-Rio), Brazil. We are also especially grateful to Dr. Ludimar Aguiar and Dr. Marcos A. Martins, from CENPES/Petrobras, for their constructive comments and useful discussions on riser design. The authors would also like to thank Dr. F\'abio Ramos for many fruitful discussions on machine learning and its applications. 

Any opinions, findings, conclusions, or recommendations expressed here are those of the authors and do not necessarily reflect the views of the sponsors.

  \bibliographystyle{plain}
  \bibliography{active_bib}

\begin{thebibliography}{10}

\bibitem{cardoso2019optimization}
PHS Cardoso, NAG Casaprima, FV~Senhora, and IFM Menezes.
\newblock Optimization of catenary risers with hydrodynamic dampers.
\newblock {\em Ocean Engineering}, 184:134--142, 2019.

\bibitem{clausen2001}
T~Clausen and R~D'Souza.
\newblock Dynamic risers key component for deepwater drilling, floating
  production.
\newblock {\em Offshore}, 61(5):89--90, 2001.

\bibitem{PhysRevLett.117.135502}
Felix~A. Faber, Alexander Lindmaa, O.~Anatole von Lilienfeld, and Rickard
  Armiento.
\newblock Machine learning energies of 2 million elpasolite
  $(ab{C}_{2}{D}_{6})$ crystals.
\newblock {\em Phys. Rev. Lett.}, 117:135502, Sep 2016.

\bibitem{DNV-OF-F201}
DNV GL.
\newblock Dnvgl-st-f201 dynamic risers, January 2018.

\bibitem{jonswap73}
Klaus Hasselmann, T.~Barnett, E.~Bouws, H.~Carlson, D.~Cartwright, K~Enke,
  J~Ewing, H~Gienapp, D.~Hasselmann, P.~Kruseman, A~Meerburg, Peter Muller,
  Dirk Olbers, K~Richter, W.~Sell, and H.~Walden.
\newblock Measurements of wind-wave growth and swell decay during the joint
  north sea wave project (jonswap).
\newblock {\em Deut. Hydrogr. Z.}, 8:1--95, 01 1973.

\bibitem{jacob1999alternative}
Breno~P Jacob, Marta~CT Reyes, Beatriz~SLP de~Lima, Ana~LFL Torres, Marcio~M
  Mourelle, Renato Silva, et~al.
\newblock Alternative configurations for steel catenary risers for
  turret-moored fpsos.
\newblock In {\em The Ninth International Offshore and Polar Engineering
  Conference}. International Society of Offshore and Polar Engineers, 1999.

\bibitem{10.5555/188490.188495}
David~D. Lewis and William~A. Gale.
\newblock A sequential algorithm for training text classifiers.
\newblock In {\em Proceedings of the 17th Annual International ACM SIGIR
  Conference on Research and Development in Information Retrieval}, SIGIR
  ’94, page 3–12, Berlin, Heidelberg, 1994. Springer-Verlag.

\bibitem{LIU2017159}
Yue Liu, Tianlu Zhao, Wangwei Ju, and Siqi Shi.
\newblock Materials discovery and design using machine learning.
\newblock {\em Journal of Materiomics}, 3(3):159 -- 177, 2017.
\newblock High-throughput Experimental and Modeling Research toward Advanced
  Batteries.

\bibitem{macKay92}
D.~J.~C. {MacKay}.
\newblock Information-based objective functions for active data selection.
\newblock {\em Neural Computation}, 4(4):590--604, July 1992.

\bibitem{McKay1979}
M.~D. McKay, R.~J. Beckman, and W.~J. Conover.
\newblock A comparison of three methods for selecting values of input variables
  in the analysis of output from a computer code.
\newblock {\em Technometrics}, 21(2):239, May 1979.

\bibitem{BLCP:CN026504553}
M.~M. Mourelle, E.~C. Gonzalez, B.~P. Jacob, and F.~L.~L. Carneiro.
\newblock Anflex - computational system for flexible and rigid riser analysis,
  international symposium; 9th, offshore engineering.
\newblock In {\em Offshore engineering, International symposium; 9th, Offshore
  engineering}, pages 441--458, Chichester, 1995. Wiley;.

\bibitem{oliphant2006guide}
Travis~E Oliphant.
\newblock {\em A guide to NumPy}, volume~1.
\newblock Trelgol Publishing USA, 2006.

\bibitem{10.1115/1.4044690}
Jitesh~H. Panchal, Mark Fuge, Ying Liu, Samy Missoum, and Conrad Tucker.
\newblock {Special Issue: Machine Learning for Engineering Design}.
\newblock {\em Journal of Mechanical Design}, 141(11), 10 2019.
\newblock 110301.

\bibitem{reback2020pandas}
The pandas~development team.
\newblock pandas-dev/pandas: Pandas, February 2020.

\bibitem{scikit-learn}
F.~Pedregosa, G.~Varoquaux, A.~Gramfort, V.~Michel, B.~Thirion, O.~Grisel,
  M.~Blondel, P.~Prettenhofer, R.~Weiss, V.~Dubourg, J.~Vanderplas, A.~Passos,
  D.~Cournapeau, M.~Brucher, M.~Perrot, and E.~Duchesnay.
\newblock Scikit-learn: Machine learning in {P}ython.
\newblock {\em Journal of Machine Learning Research}, 12:2825--2830, 2011.

\bibitem{PER-GRA:2007}
Fernando P\'erez and Brian~E. Granger.
\newblock {IP}ython: a system for interactive scientific computing.
\newblock {\em Computing in Science and Engineering}, 9(3):21--29, May 2007.

\bibitem{doi:10.1063/1.4977912}
J.~L. Peterson, K.~D. Humbird, J.~E. Field, S.~T. Brandon, S.~H. Langer, R.~C.
  Nora, B.~K. Spears, and P.~T. Springer.
\newblock Zonal flow generation in inertial confinement fusion implosions.
\newblock {\em Physics of Plasmas}, 24(3):032702, 2017.

\bibitem{Rasmussen:2005:GPM:1162254}
Carl~Edward Rasmussen and Christopher K.~I. Williams.
\newblock {\em Gaussian Processes for Machine Learning (Adaptive Computation
  and Machine Learning)}.
\newblock The MIT Press, 2005.

\bibitem{settles.tr09}
Burr Settles.
\newblock Active learning literature survey.
\newblock Computer Sciences Technical Report 1648, University of
  Wisconsin--Madison, 2009.

\bibitem{shannonEntropy}
C.~E. {Shannon}.
\newblock A mathematical theory of communication.
\newblock {\em The Bell System Technical Journal}, 27(3):379--423, July 1948.

\bibitem{10.1115/1.2429697}
G.~Gary Wang and S.~Shan.
\newblock {Review of Metamodeling Techniques in Support of Engineering Design
  Optimization}.
\newblock {\em Journal of Mechanical Design}, 129(4):370--380, 05 2006.

\bibitem{mckinney-proc-scipy-2010}
{W}es {M}c{K}inney.
\newblock {D}ata {S}tructures for {S}tatistical {C}omputing in {P}ython.
\newblock In {S}t\'efan van~der {W}alt and {J}arrod {M}illman, editors, {\em
  {P}roceedings of the 9th {P}ython in {S}cience {C}onference}, pages 56 -- 61,
  2010.

\end{thebibliography}

\end{document}